\documentclass[usenatbib]{mn2e}
\usepackage{epsfig}
\usepackage{amsmath}
\usepackage{ulem}
\usepackage{times}
\usepackage{longtable}
\usepackage{supertabular}
\usepackage{color}

\def\gsim{\mathrel{\raise0.35ex\hbox{$\scriptstyle >$}\kern-0.6em 
\lower0.40ex\hbox{{$\scriptstyle \sim$}}}}
\def\lsim{\mathrel{\raise0.35ex\hbox{$\scriptstyle <$}\kern-0.6em 
\lower0.40ex\hbox{{$\scriptstyle \sim$}}}}

\date{\today}
\title[The SFR v.s. $M_*$ relation since $z\sim 2$]
{On the evolution and environmental dependence of the star formation rate versus stellar mass relation since {\boldmath{$z\sim 2$}}}

\author[Y. Koyama et al.]{
\parbox[t]{\textwidth}{
Yusei Koyama,$^{\! 1,2}$\thanks{E-mail: yusei.koyama@durham.ac.uk}
Ian Smail,$^{\! 3}$
Jaron Kurk,$^{\! 4}$
James E. Geach,$^{\! 5,6}$
David Sobral,$^{\! 7}$\\
Tadayuki Kodama,$^{\! 2,8}$
Fumiaki Nakata,$^{\! 2,8}$
A. M. Swinbank,$^{\! 3}$
Philip N. Best,$^{\! 9}$\\
Masao Hayashi,$^{\! 2,10}$
Ken-ichi Tadaki$^{11}$
}
\vspace*{6pt}\\
$^{1}$Department of Physics, Durham University, South Road, Durham DH1 3LE, UK\\
$^{2}$National Astronomical Observatory of Japan, Mitaka, Tokyo
181-8588, Japan\\
$^{3}$Institute for Computational Cosmology, Durham University, South Road, Durham DH1 3LE, UK\\
$^{4}$Max-Planck-Institut f{\"u}r Extraterrestrische Physik, Postfach 1312, Giessenbachstrasse, D-85741 Garching, Germany\\
$^{5}$Department of Physics, McGill University, 3600 Rue University Montr\'{e}al, Qu\'{e}bec H3A 2T8, Canada\\
$^{6}$Centre for Astrophysics Research, Science \& Technology Research Institute, University of Hertfordshire, Hatfield, AL10 9AB\\
$^{7}$Leiden Observatory, Leiden University, PO Box 9513, NL-2300 RA Leiden, the Netherlands\\
$^{8}$Subaru Telescope, National Astronomical Observatory of Japan, 650, North A'ohoku Place, Hilo, HI 96720, USA\\
$^{9}$SUPA, Institute for Astronomy, Royal Observatory of Edinburgh, Blackford Hill, Edinburgh EH9 3HJ\\
$^{10}$Institute for Cosmic Ray Research, The University of Tokyo, Kashiwa, 277-8582, Japan\\
$^{11}$Department of Astronomy, Graduate School of Science, The University of Tokyo, Tokyo 113-0033, Japan 
}
\begin{document}

\maketitle

\begin{abstract}
This paper discusses the evolution of the correlation between galaxy star formation rates (SFRs) and stellar mass ($M_*$) over the last $\sim$10 Gyrs, particularly focusing on its environmental dependence. We first present the mid-infrared (MIR) properties of the H$\alpha$-selected galaxies in a rich cluster Cl\,0939+4713 at $z=0.4$. We use wide-field Spitzer/MIPS24$\mu$m data to show that the optically red H$\alpha$ emitters, which are most prevalent in group-scale environments, tend to have higher SFRs and higher dust extinction than the majority population of blue H$\alpha$ sources. With a MIR stacking analysis, we find that the median SFR of H$\alpha$ emitters is higher in higher-density environment at $z=0.4$. We also find that star-forming galaxies in high-density environment tend to have higher specific SFR (SSFR), although the trend is much less significant compared to that of SFR. This increase of SSFR in high-density environment is not visible when we consider the SFR derived from H$\alpha$ alone, suggesting that the dust attenuation in galaxies depends on environment; galaxies in high-density environment tend to be dustier (by up to $\sim$0.5 mag), probably reflecting a higher fraction of nucleated, dusty star-bursts in higher-density environments at $z=0.4$. We then discuss the environmental dependence of the SFR--$M_*$ relation for star-forming galaxies since $z\sim 2$, by compiling our comparable, narrow-band selected, large H$\alpha$ emitter samples in both distant cluster environments (from MAHALO-Subaru) and field environments (from HiZELS). We find that the SSFR of H$\alpha$-selected galaxies (at the fixed mass of $\log (M_*/M_{\odot})=$10) rapidly evolves as $(1+z)^3$, but the SFR--$M_*$ relation is independent of the environment since $z\sim 2$, as far as we rely on the H$\alpha$-based SFRs (with $M_*$-dependent extinction correction). Even if we consider the possible environmental variation in the dust attenuation, we conclude that the difference in the SFR--$M_*$ relation between cluster and field star-forming galaxies is always small ($\lsim$0.2~dex level) at any time in the history of the Universe since $z\sim 2$.

\end{abstract}
\section{Introduction}
\label{sec:intro}

Galaxy formation and evolution is strongly dependent on environment. In the local Universe, galaxies in cluster environments are mostly passive (red), early-type galaxies (e.g.\ \citealt{dre80}; \citealt{got03}), and there is a clear trend that the star formation activity of galaxies tends to be lower in high-density environment than low-density fields (e.g.\ \citealt{lew02}; \citealt{gom03}; \citealt{tan04}; \citealt{bal04}). Therefore it is believed that the star formation activity of galaxies is affected by their surrounding environment during the course of the cluster or group assembly process. The environmental trends are also seen in the distant Universe (e.g.\ \citealt{kod01}; \citealt{pos05}; \citealt{qua12}), and some recent studies suggest that the star formation--density relation may be reversed at $z\gsim 1$ (\citealt{elb07}; \citealt{coo08}; \citealt{ide09}; \citealt{tra10}). This "reversal" of the star formation--density relation in the early Universe is still under debate (\citealt{pat09}; 2011), probably reflecting the fact that the results could be uncertain depending on the sample definitions or the definitions of environment (e.g.\ \citealt{fer10}; \citealt{pop11}; \citealt{sob11}). Nevertheless, some recent studies which focus on individual galaxy clusters indeed find a hint that a fraction of galaxies in distant ($z\sim 1$) cluster outskirts or intermediate-density environments are showing boosted activity (e.g.\ \citealt{mar07}; \citealt{pog08}; 2009; \citealt{koy10}; \citealt{gea11}), suggesting an accelerated galaxy evolution at the site of active cluster assembly.

A growing number of studies have revealed important roles of the group-scale environment or in-falling regions around rich clusters, by studying e.g. galaxy colours or morphologies (e.g.\ \citealt{kod01}; \citealt{wil08}; \citealt{bal11}). In addition, optical emission-line surveys or MIR--FIR observations of distant clusters have brought some new insights into the (obscured) nature of star forming galaxies in cluster environment. A prominent example are {\it dusty red galaxies}, which are reported to populate the outskirts of rich galaxy clusters out to $z\sim 1$ (e.g.\ \citealt{koy08}; 2010; see also \citealt{wol05}; \citealt{gea06}; \citealt{ver08}; \citealt{wol09}; \citealt{tra09}). These studies suggest that such dusty red galaxies are a key population for understanding the physics of environmental effects. In particular, they are recognised as the strong candidates for the progenitors of local cluster S0 galaxies (e.g.\ \citealt{gea09}; \citealt{koy11}), in the phase of rapid 'bulge growth' which is required to explain the rapid increase of the S0 galaxy fraction in clusters since $z\sim 1$ (e.g.\ \citealt{dre97}; \citealt{kod01}). 

In this respect, wide-field H$\alpha$$\lambda$6563 emission-line surveys of distant galaxy clusters are a powerful method to pinpoint the location of this key population. This is not only because the H$\alpha$ line is less affected by dust extinction compared to star-formation indicators at rest-frame ultra-violet wavelengths (see e.g.\ \cite{hay13} showing [{\sc Oii}]$\lambda$3727 only recovers relatively 'dust-free' population), but also because emission line surveys with narrow-band (NB) filters allow us to effectively pick out galaxies from a narrow redshift slice. The latter is particularly important for cluster studies because the effect of contamination could be a concern. \cite{koy11} performed a wide-field H$\alpha$ emission-line survey of a rich cluster, Cl\,0939+4713 ($z=0.41$), and find a strong concentration of optically red star-forming galaxies in the group-scale environment around the cluster. While we argued in \cite{koy11} that the excess of the red star-forming galaxies suggests an enhancement of dust-obscured star formation in the group environment, a firm conclusion still await a direct measurement of dust-enshrouded star-formation in those galaxies, because even H$\alpha$ lines are reported to be heavily extinguished in extremely dusty galaxies (e.g.\ \citealt{pog00}; \citealt{gea06}; \citealt{koy10}). Therefore, the first goal of this paper is to directly unveil the nature of this red star-forming population in distant group environments using mid-infrared (MIR) observations.

Another important parameter that drives galaxy evolution is the stellar mass ($M_*$) of galaxies. A correlation has been claimed between galaxy star-formation rate (SFR) and $M_*$ for star-forming galaxies in the local Universe (e.g.\ \citealt{bri04}; \citealt{pen10}), as well as in the distant Universe out to $z\gsim 2$ (e.g.\ \citealt{noe07}; \citealt{dad07}; \citealt{san09}; \citealt{kaj10}; \citealt{bau11}; \citealt{whi12}). This correlation is often called the "Main Sequence" of star forming galaxies, and a growing number of studies are now investigating various aspects of this relation; e.g. the origin of its scatter or its morphological dependence (\citealt{wuy11}; \citealt{sal12}). An interesting implication from some detailed studies of the local SFR--$M_*$ relation is the "independence" of this relation with environment; \cite{pen10} studied local star-forming galaxies drawn from SDSS to show an excellent agreement in the SFR--$M_*$ sequence between low-density and high-density environments. They argue that the environment does change the star-forming galaxy {\it fraction}, but that it has very little impact on the SFR--$M_*$ relation of those galaxies that are star-forming (see also \citealt{wij12}).

An observational challenge is to test this universality of SFR--$M_*$ relation in the distant Universe. Any local relation may not necessarily be applicable for distant galaxies at $z\gsim 1$, where the average star-formation activity is about an order of magnitude higher (e.g.\ \citealt{hop06}; see also \citealt{sob13}; \citealt{sto13}). Unfortunately, constructing a large, uniformly selected galaxy sample at such high redshifts is still challenging, which prohibits us from understanding the environmental impacts on the SFR--$M_*$ relation in the distant Universe. Some earlier works have attempted to identify the environmental dependence (or its absence) of the SFR--$M_*$ relation out to $z=1$ (\citealt{vul10}; \citealt{li11}; \citealt{mcg11}; \citealt{tyl11}; \citealt{muz12}), or out to $z=2$ (\citealt{tan10}; \citealt{tan11}; \citealt{gru11}; \citealt{koy13}), but a full consensus on the environmental impacts on the SFR versus $M_*$ relation has not yet been reached, because of the different sample selection and/or different environment definitions. Therefore, the second goal of this paper is to test the environmental dependence of the SFR--$M_*$ relation out to $z\sim 2$ for the first time based on the purely H$\alpha$-selected star-forming galaxy samples established in our recent two narrow-band H$\alpha$ survey projects; MAHALO-Subaru for clusters and HiZELS for general fields (see details in \S~2.2).

The structure of this paper is as follows. In \S~2, we present the wide-field Subaru and {\it Spitzer}/MIPS 24$\mu$m data of Cl\,0939, as well as the H$\alpha$ emitter samples at different redshifts and in different environments from our previous studies. We also describe how we estimate fundamental physical quantities such as stellar masses ($M_*$) and star formation rates (SFR) of galaxies. The main results and discussion of the paper are described in \S~3. We first discuss MIR properties of H$\alpha$ emitters around Cl\,0939, particularly focusing on the nature of the red H$\alpha$ sources (\S~3.1--\S~3.2), and then we discuss the environmental dependence of the SFR versus $M_*$ relation since $z\sim 2$ using all the H$\alpha$ selected galaxies (\S~3.3--\S~3.6). Finally, our conclusions are given in \S~4. Throughout the paper, we adopt the standard cosmology with $\Omega_M =0.3$, $\Omega_{\Lambda} =0.7$, and $H_0=70$\,km\,s$^{-1}$\,Mpc$^{-1}$, and we assume a \cite{sal55} initial mass function (IMF). All physical quantities (SFR and $M_*$) derived in the literatures assuming different IMFs are rescaled to the Salpeter IMF. Magnitudes are all given in the AB system.

\section{Data and Analysis}

\subsection{The Cl\,0939 cluster}

\subsubsection{Subaru data}

The Cl\,0939 cluster at $z=0.41$ is one of the best-studied clusters at intermediate redshifts (e.g.\ \citealt{dre92}; \citealt{dre94}; \citealt{sta95}; \citealt{sma99}; \citealt{sat06}; \citealt{dre09}; \citealt{oem09}). In addition to these studies which focus on cluster central regions, very wide-field ($\sim$30$'\times$30$'$) optical broad-band and narrow-band imaging surveys of this cluster have been made by \cite{kod01} and \cite{koy11} using Suprime-Cam (\citealt{miy02}) on the Subaru Telescope (\citealt{iye04}). The details of these Subaru data have already been presented in our previous papers (\citealt{kod01}; \citealt{koy11}), and so here we provide only a quick summary of the data. The broad-band data are analysed by \cite{kod01}. They discovered a 10~Mpc-scale filamentary large-scale structure around the cluster based on the photometric redshift (photo-$z$) technique. The H$\alpha$ emitter search of this field is made by \cite{koy11}, using the narrow-band filter NB921 ($\lambda_c$$=$9180\AA) on Suprime-Cam, who identified $>$400 H$\alpha$ emitting galaxies around the cluster. In this paper, we use both the H$\alpha$ emitter and the photo-$z$ selected cluster member catalogues presented in these studies. The photometry of the sources is performed with the {\sc SExtractor} software (\citealt{ber96}). We primarily use MAG\_AUTO as the total magnitudes for measuring physical quantities such as $M_*$ or SFR, while we use 3$''$ aperture (corresponding to 16~kpc) photometry for measuring galaxy colours.

\subsubsection{Spitzer MIPS 24$\mu$m data}

We retrieve the wide-field MIPS (Rieke et al. 2004) 24$\mu$m scan data of the Cl\,0939 field from the {\it Spitzer} Science Archive. The data covers a large part of our Suprime-Cam field of view (from cluster core to surrounding groups), so are well-suited for studying dust-obscured star formation activity around the cluster. The data was reduced from the Basic Calibrated Data stage (BCD; provided by the {\it Spitzer} Science Centre) using the MOPEX software following the procedure outlined in Geach et al. (2006). Source extraction was performed using SExtractor (\citealt{ber96}), with the criteria that a source consist of at least three contiguous pixels (each pixel is 2.5$''$ square) at $>$2$\sigma$ above the background. We measure 16$''$ diameter aperture fluxes, corresponding to $\sim$3 times the FWHM of the point-spread function (PSF; FWHM of ~5$''$ at 24$\mu$m). Using a curve-of-growth analysis on bright isolated point sources, and our completeness simulations, we find that 16$''$ apertures recover $\sim$75 per cent of the total flux, and we therefore correct the resulting fluxes by a factor of 1.33 to yield the total 24$\mu$m fluxes. 

The 24$\mu$m catalogue contains 886 sources down to 200$\mu$Jy ($\sim$5$\sigma$ limit) within the Suprime-Cam field. We perform cross-matching between the 24$\mu$m sources and our optical sources to construct a MIPS-detected cluster member catalogue. We search for optical sources within 3$''$ radius from each 24$\mu$m source, and we select the nearest source as its optical counterpart. As a result, we find 162 MIPS sources which are likely to be associated with cluster member galaxies (photo-$z$ members or H$\alpha$ emitters), although due to the poor spatial resolution of the MIPS24$\mu$m data, it is sometimes difficult to identify the counterparts correctly. In order to avoid such ambiguous detections, we check all the sources by eye and exclude some heavily blended sources. Only 10 sources (6\%) are excluded in this process (and they seem to be just normal galaxies randomly distributed on sky), so the exclusion of these sources does not affect our results at all. Overall, the MIPS-detected member catalogue contains total 152 sources, among which 33 sources are H$\alpha$ emitters. The relatively small number of the H$\alpha$ detected sources is not surprising, because the photo-$z$ selected members include galaxies over a relatively broad range in redshift (we apply 0.30$\le$$z_{\rm phot}$$\le$0.45 for photo-$z$ member criteria; see \citealt{koy11}), while the H$\alpha$ emitters are considered to be secure cluster members located within the narrow redshift slice at the cluster's redshift (0.39$\le$$z$$\le$0.41).

\subsubsection{MIPS stacking analysis}

We note that the limiting flux of the 24$\mu$m data corresponds to SFR$\sim$4$M_{\odot}$\,yr$^{-1}$\footnote{This is a typical SFR limit derived from the H$\alpha$ and $L_{\rm IR}$ approach (see Fig.~2). The limiting SFR will increase by a factor of $\sim$2$\times$ if we directly convert $L_{\rm IR}$ to SFR.}, while our H$\alpha$ survey reaches down to SFR$\ll$1$M_{\odot}$\,yr$^{-1}$ (without dust extinction correction). Indeed, the number of H$\alpha$ emitters which are individually detected at 24$\mu$m is not large, because of the limited depths of the 24$\mu$m data.  We therefore apply a 24$\mu$m stacking analysis for the H$\alpha$ emitters around Cl\,0939 to study the general properties of the fainter sources, by dividing the sample into subsamples selected by colour or environment (see \S~3.2, \S~3.3). We exploit median stacking at the positions of H$\alpha$ emitters down to an dust-uncorrected SFR of 0.25$M_{\odot}$\,yr$^{-1}$ (which corresponds to a 5$\sigma$ detection in H$\alpha$). Although our conclusion does not change even if we apply an average stacking, median stacking is preferred because it can minimize the effects from exceptionally luminous sources, as well as those from some luminous nearby sources. The latter could be a concern particularly for high-density cluster environments, but we note that the MIPS source density is not significantly higher in the cluster region (reflecting the fact that most galaxies in the cluster core are not star-forming; \citealt{koy11}), so that it does not affect our results.

The photometry on the MIPS stacked image is performed in the same way as for individual sources; i.e.\ 16$''$ aperture photometry with the aperture correction of $\times$1.33 (\S~2.1.2). Also, we apply bootstrap re-sampling (500 times) to obtain the 1-$\sigma$ error-bars for the median 24$\mu$m fluxes from stacking. Note that we include all the H$\alpha$ emitters for the stacking, regardless of their individual 24$\mu$m detection, while we do not use the sources without H$\alpha$ detection in order to avoid contaminant galaxies. We derive the total infrared luminosity ($L_{\rm IR}$) from the stacked 24$\mu$m photometry using the \cite{cha01} SED templates, and then compute the SFR using the combined H$\alpha$ and $L_{\rm IR}$ approach suggested by \cite{ken09}; SFR$=7.9\times 10^{-42}[L$(H$\alpha$)$_{\rm  obs}+0.0024\times L_{\rm IR}]$~(erg\,s$^{-1}$). This equation is derived from a tight correlation between the combined H$\alpha$ and total IR luminosities and the attenuation corrected SFR (see \citealt{ken09}). We note that our conclusion does not change even if we calculate SFRs directly from the total IR luminosity using the \cite{ken98} relation (in this case we tend to derive SFRs which are higher by a factor of $\lsim$2), but the combined H$\alpha$ and IR approach should be more reliable, considering the moderate levels of star-formation activity (with $\lsim$10$M_{\odot}$\,yr$^{-1}$) of our galaxy sample.

\subsection{H$\alpha$ emitter samples from our previous studies}

\subsubsection{Cluster galaxy sample from MAHALO-Subaru}

In addition to the Cl\,0939 cluster ($z=0.4$), we also use our similar, H$\alpha$ selected galaxies in a $z=0.8$ cluster (RXJ1716+6708) and in a $z=2.2$ proto-cluster (PKS1138$-$262), from our previous studies (see \citealt{koy10}; 2013 for detailed descriptions of these data and the selection of H$\alpha$ emitters). The H$\alpha$ emitter samples are constructed with narrow-band surveys using NB119 ($\lambda_c=1.19\mu$m for $z=0.8$) and NB2071 ($\lambda_c=2.07\mu$m for $z=2.2$) on MOIRCS/Subaru (\citealt{ich06}; \citealt{suz08}), as a part of the {\it MApping H-Alpha and Lines of Oxygen with Subaru} project (MAHALO-Subaru; see overview by \citealt{kod13}). We use these cluster galaxy samples to study the redshift evolution of the SFR--$M_*$ relation in cluster environments (see \S~3.4). We note that our advantage is a perfect matching of their redshifts with our similar H$\alpha$-selected control field galaxy samples from HiZELS (see below). Using only one cluster at each redshift might be too simplistic (e.g.\ our proto-cluster at $z=2.2$ may not necessarily be the progenitor of our $z=0.4$ and $0.8$ clusters), but in this pioneering study, we assume that these H$\alpha$ emitter samples in these three cluster fields represent typical star forming galaxies in high-density environments at each redshift.

 \begin{figure}
  \begin{center}
    \leavevmode
    \vspace{-0.2cm}
    \rotatebox{0}{\includegraphics[width=8.4cm,height=8.4cm]{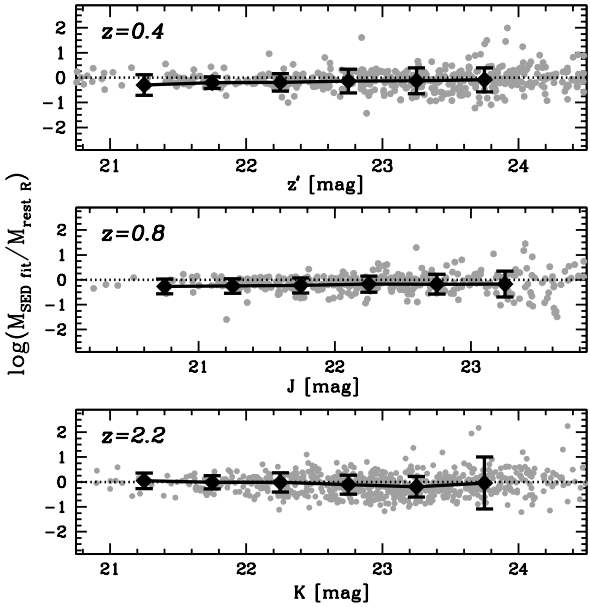}}
  \end{center}
  \vspace{-0.3cm}
  \caption{ A comparison between the stellar mass estimates for HiZELS
    H$\alpha$ emitters derived from the rest-frame {\it R}-band
    magnitudes with one-colour correction (see text) and those from
    full SED fitting derived by Sobral et al. (in prep). The
    line-connected black points show the running median (and its
    associated scatter). It is clear that these two measurements
    agree with each other reasonably well over a wide luminosity range. }
  \label{fig:mstar_check}
\end{figure}

\subsubsection{Field galaxy sample from HiZELS}

In order to test the environmental dependence of the SFR versus $M_*$ relation, we also need a control field galaxy sample. The {\it High-Z Emission Line Survey} (HiZELS; \citealt{bes10}; \citealt{sob13}) is ideally suited for this purpose. This is not only because HiZELS is currently the largest narrow-band H$\alpha$ survey ever published, but also because three of their four targeted redshifts ($z=0.4/0.8/2.2$) are perfectly matched with our cluster samples, allowing a direct cluster--field comparison based on the purely H$\alpha$-selected galaxies. The HiZELS H$\alpha$ emitter samples are selected from the UDS and COSMOS fields (total $\sim$2~deg$^2$; see also \citealt{gea08}; 2012; \citealt{sob09}; 2012; 2013) by wide-field narrow-band imaging observations with NB921 on Subaru ($z=0.4$), NB$_{J}$ on UKIRT ($z=0.8$), and the NB$_{K}$/H$_2$ filters on UKIRT/VLT ($z=2.2$). We select H$\alpha$ emitters in exactly the same way as described in \cite{sob13}. We note that the selection and photometry of the HiZELS sources has been made with 3$''$ and 2$''$ aperture for $z=0.4$ and $z=0.8$/$2.2$ galaxies (see \citealt{sob13}), respectively, while physical quantities of our cluster galaxy samples are measured with total magnitudes ({\sc mag\_auto} from {\sc SExtractor}). Therefore, we apply an aperture correction for the HiZELS sample based on the median difference between aperture magnitudes and total magnitudes for each redshift slice, although these corrections are negligibly small in our discussion (0.2~dex at maximum).

\subsection{Stellar mass and H$\alpha$-based star formation rate}

In this subsection, we derive stellar masses ($M_*$) and star formation rates (SFR) of the H$\alpha$ emitters. The stellar masses of galaxies are ideally derived using a SED fitting approach including rest-frame near-infrared (NIR) photometry. However, the rest-frame NIR photometry is not available for our cluster galaxy samples. We therefore decide to estimate the stellar masses of galaxies based on their rest-frame {\it R}-band magnitudes. The conversions from observed magnitudes to $M_*$ are determined using the model galaxies developed by \cite{kod99} (see also \citealt{kod97}; \citealt{kod98}), and they are expressed by the following equations:
\begin{equation}
\log (M_*/10^{11}M_{\odot})_{z=0.4} = -0.4(z'-20.07) + \Delta\log M_{0.4},
\end{equation}
\begin{equation}
\log (M_*/10^{11}M_{\odot})_{z=0.8} = -0.4(J-21.14) + \Delta\log M_{0.8},
\end{equation}
\begin{equation}
\log (M_*/10^{11}M_{\odot})_{z=2.2} = -0.4(K_{(s)}-22.24) + \Delta\log M_{2.2}.
\end{equation}
The final term in each equation ($\Delta M$) accounts for the colour dependence of the mass-to-light ratio ($M/L$) predicted by the same galaxy model, and they correspond to:
\begin{equation}
\Delta\log M_{0.4}= 0.054 - 3.81\times \exp[-1.28\times (B-z')],
\end{equation}
\begin{equation}
\Delta\log M_{0.8}= 0.085 - 2.48\times \exp[-1.29\times (R-J)],
\end{equation}
\begin{equation}
\Delta\log M_{2.2}= 0.030 - 1.50\times \exp[-1.11\times (z'-K_{(s)})].
\end{equation}
The stellar masses derived with this ``one-colour method'' could be less accurate compared to those from full SED fitting, but we find that there is no systematic difference between the $M_*$ from this method and those from the SED-fitting approach using HiZELS H$\alpha$ emitter samples for which full SED information are available (see Fig.~\ref{fig:mstar_check}). The two measurements are consistent with each other over a wide luminosity range (with a $\approx$0.3~dex scatter), verifying that our one-colour method works reasonably well. In this paper, in order to make a fair comparison between cluster and field samples at different redshifts, we use the above one-colour method for all H$\alpha$ emitters (including MAHALO and HiZELS samples).

 \begin{figure}
  \begin{center}
    \leavevmode
    \vspace{-0.6cm}
    \rotatebox{0}{\includegraphics[width=8.5cm,height=8.5cm]{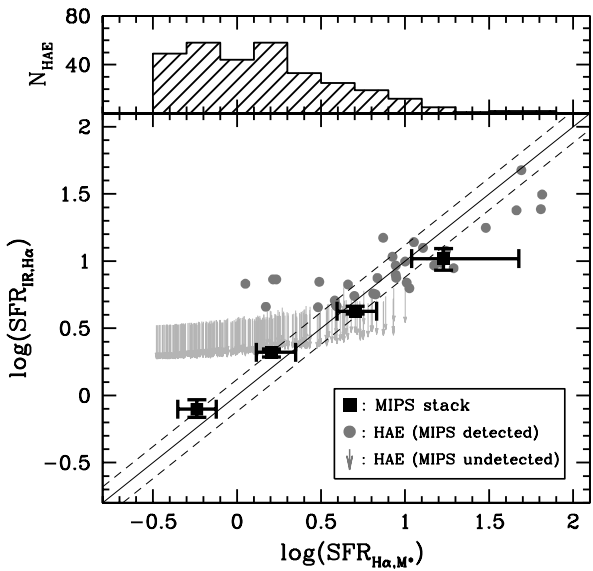}}
  \end{center}
  \vspace{-0.7cm}
  \caption{ A comparison between the SFR$_{\rm (IR,H\alpha)}$ and the extinction-corrected SFR$_{\rm (H\alpha,M_*)}$. In addition to the individual data points (grey symbols), we also apply the MIPS stacking analysis by dividing the sample at $\log$(SFR$_{\rm H\alpha,M_*}$)$=0.0,0.5,1.0$ (see black squares with error-bars). The vertical error-bars show the 1-$\sigma$ uncertainty derived from the bootstrap re-sampling approach during the MIPS stacking (see text), while the horizontal error-bars show the 25--75\% distribution of each subsample. The solid and dashed lines show the one-to-one relation with a typical uncertainty in the extinction correction ($\sim$0.3 mag), as reported in Garn \& Best (2010). The entire distribution of SFR$_{\rm (H\alpha,M_*)}$ is shown in the top panel.
}
  \label{fig:sfr_check}
\end{figure}

We then calculate SFRs of H$\alpha$ emitters based on their H$\alpha$ luminosities. We first correct for the contribution of [{\sc Nii}] lines to the total NB fluxes, using an empirical correlation between EW$_{\rm rest}$(H$\alpha$+[{\sc Nii}]) and the [{\sc Nii}]/H$\alpha$ ratio for local galaxies as described in \cite{sob13}. This relation has a large intrinsic scatter, but this method is believed to be more realistic approach compared with the conventional constant [{\sc Nii}] correction. We also apply a dust extinction correction to the H$\alpha$ flux of individual galaxies based on their stellar mass, as shown by \cite{gar10b}. This extinction correction could also be uncertain, given the large intrinsic scatter of the $A_{\rm H\alpha}$--$M_*$ relation for local galaxies (see \citealt{gar10b}). However, it is one of the most reliable estimators of dust extinction, and also the relation is reported to be unchanged out to $z\sim 1.5$ (\citealt{sob12}; see also \citealt{dom13}, \citealt{iba13}). We therefore apply the same extinction correction to all H$\alpha$ emitters. 

Finally, we compute the SFRs of galaxies using the \cite{ken98} relation; SFR($M_{\odot}$\,yr$^{-1}$)$=7.9\times10^{-42}L_{\rm{H}\alpha}$(erg\,s$^{-1}$). We note again that the [{\sc Nii}] line and the dust extinction corrections could be major sources of uncertainty, but this is currently inevitable because it is impossible to measure [{\sc Nii}] contribution or dust extinction for individual galaxies. As a quick check, we compare in Fig.~\ref{fig:sfr_check} the SFRs from H$\alpha$ with $M_*$-dependent extinction correction (SFR$_{{\rm (H\alpha}, M_*)}$ hereafter) with the SFRs derived from IR and H$\alpha$ approach as described in \S~2.1.3 (SFR$_{\rm (IR,H\alpha)}$ hereafter). The number of H$\alpha$ emitters individually detected at 24$\mu$m is small (only 10\% of the total sample), and so we can only show the upper limits for most galaxies (see arrows in Fig.~\ref{fig:sfr_check}). In this diagram, we can see a few sources with SFR$_{\rm (IR,H\alpha)}$$>$SFR$_{\rm (H\alpha,M_*)}$. They are probably IR luminous starbursts (with strong dust attenuation at H$\alpha$). We can also see some sources with SFR$_{\rm (H\alpha,M_*)}$$>$SFR$_{\rm (IR,H\alpha)}$ at high SFR end. This is probably because of the AGN contribution, but such galaxies (with SFR$>$10$M_{\odot}$\,yr$^{-1}$) are only $\sim$5\% in our total sample (see histogram in Fig.~\ref{fig:sfr_check}), and so the effect of such extreme sources to our main conclusion is small. By applying the MIPS stacking analysis, we find that the two SFRs agree reasonably well over a wide luminosity range (see black symbols in Fig.~\ref{fig:sfr_check}). We note that in the latter half of this paper we use the H$\alpha$-derived SFR (SFR$_{{\rm (H\alpha}, M_*)}$) to discuss the evolution and environmental dependence of star-formation activity of galaxies. We should note that using the different SFR indicators could bring slightly different results, reflecting the fact that the dust extinction could be dependent on environment (see \S~3.3). However, we do not expect it to significantly influence our conclusion, particularly for the evolutionary trend we see, because the evolutionary trend is so strong that the environmental variation will not over-ride it (\S~3.4). 

  \begin{figure*}
   \vspace{-0.4cm}
   \begin{center}
    \leavevmode
    \epsfxsize 0.48\hsize
    \epsfbox{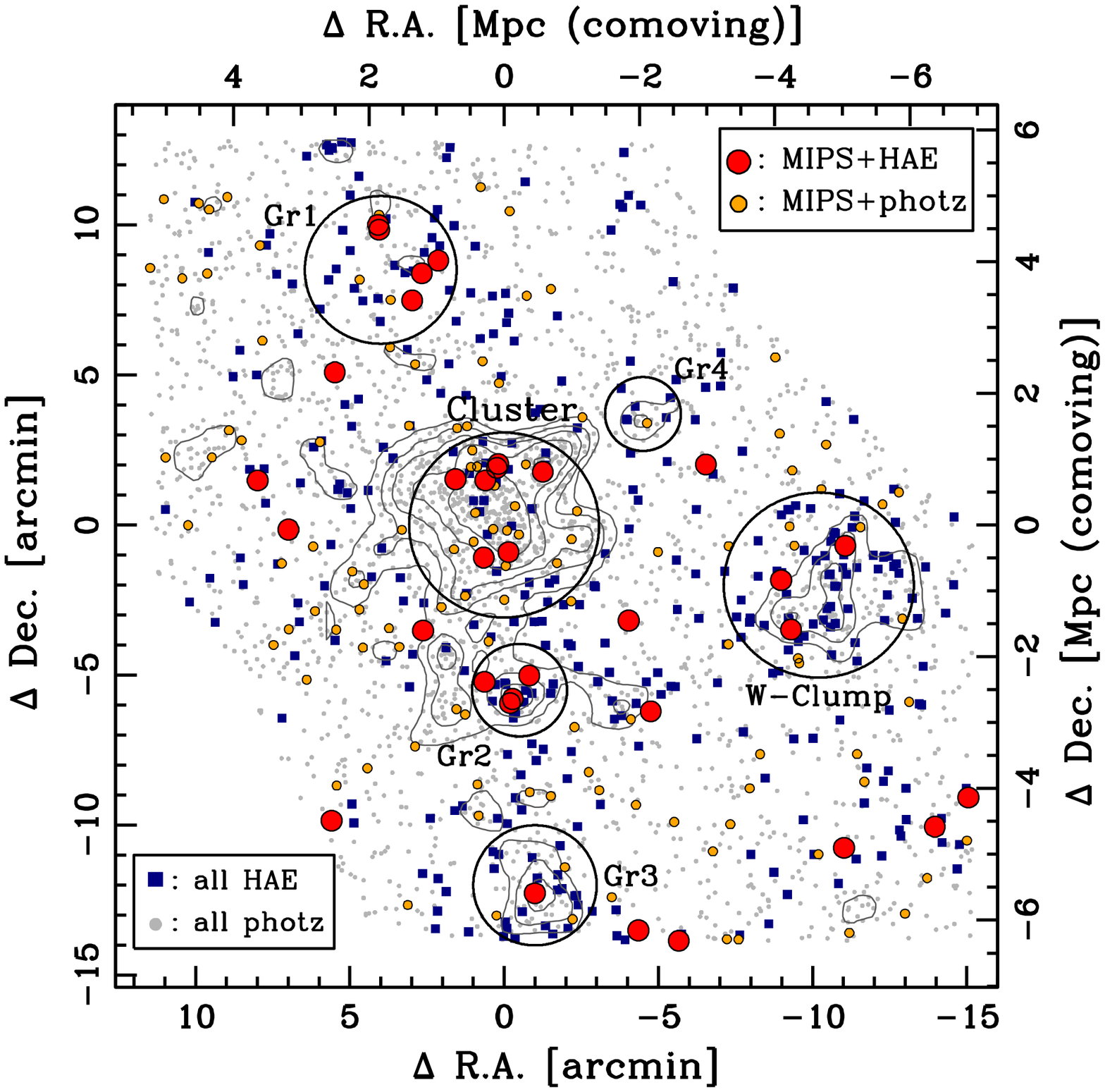}
    \hspace{-2mm}
    \epsfxsize 0.46\hsize
    \epsfbox{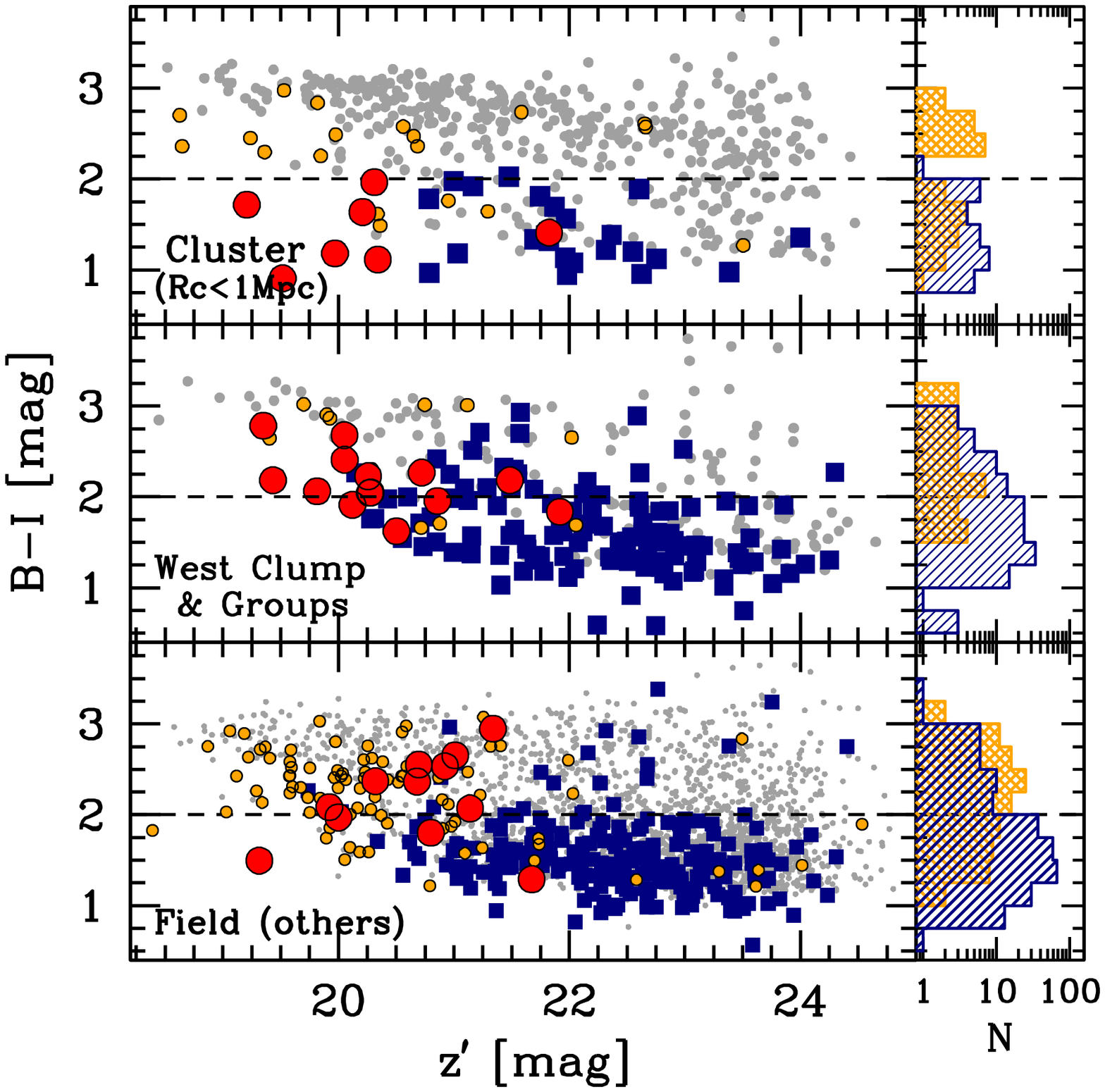}
   \end{center} 
  \vspace{-5mm}
   \caption{({\it Left}): The 2-D map of galaxies around the cluster
     Cl\,0939 ($z=0.41$). Galaxies within both Suprime-Cam and MIPS
     FoVs are shown. We plot 24$\mu$m-detected H$\alpha$ emitters,
     24$\mu$m-detected photo-$z$ members (0.30$\le$$z_{\rm
       phot}$$\le$0.45), all H$\alpha$ emitters, and all photo-$z$
     members (the meanings of the symbols are shown in the plot). We
     also show the locations of the surrounding groups following the
     definitions in Koyama et al. (2011). Contours are drawn based on
     the surface number density of all member galaxies (same as fig.~5
     of Koyama et al. 2011). ({\it Right}): The colour--magnitude
     diagram for each environment as defined in the left-hand
     panel. The meanings of the symbols are the same as the left-hand
     panel. The (blue) hatched and (orange) cross-hatched histograms
     show the colour distribution of H$\alpha$ emitters and 24$\mu$m
     sources, respectively. }
\label{fig:map_and_cmd}
 \end{figure*}

The stellar masses and SFRs derived above may not be very accurate for galaxies hosting active galactic nuclei (AGN), as their continuum light and H$\alpha$ line fluxes are contaminated by AGNs. For the field H$\alpha$ emitters (HiZELS), \cite{gar10a} and \cite{sob13} carried out a detailed study on the contribution of AGNs into their H$\alpha$ emitters sample. By applying various techniques for AGN selection (e.g.\ X-ray, radio, or emission-line ratios), they find that the AGN fraction amongst their H$\alpha$ emitters is as large as 10\% out to $z\sim 1$ or $\sim$15\% at $z>1$. For cluster samples, we confirmed a few (up to five) AGNs in each cluster using X-ray imaging data or spectroscopic information (see \citealt{koy08}; 2011; 2013). Considering the size of the total galaxy sample we use in this paper ($\sim$100 galaxies per cluster), it is roughly estimated that the AGN fraction in our cluster galaxies is $\approx$5--10\%. Unfortunately, it is not possible to fully quantify the contribution of AGNs in our cluster H$\alpha$ emitters in the same way as for HiZELS samples (due to the lack of multi-wavelength data). In order to do a fair comparison between cluster and field samples, we do not exclude AGNs from our cluster or field H$\alpha$ emitters in the following discussions, but we note again that the AGN contribution is always small ($\approx$10\%) and should not strongly bias the results.

  \begin{figure*}
   \vspace{-4mm}
   \begin{center}
    \leavevmode
    \epsfxsize 0.36\hsize
    \epsfbox{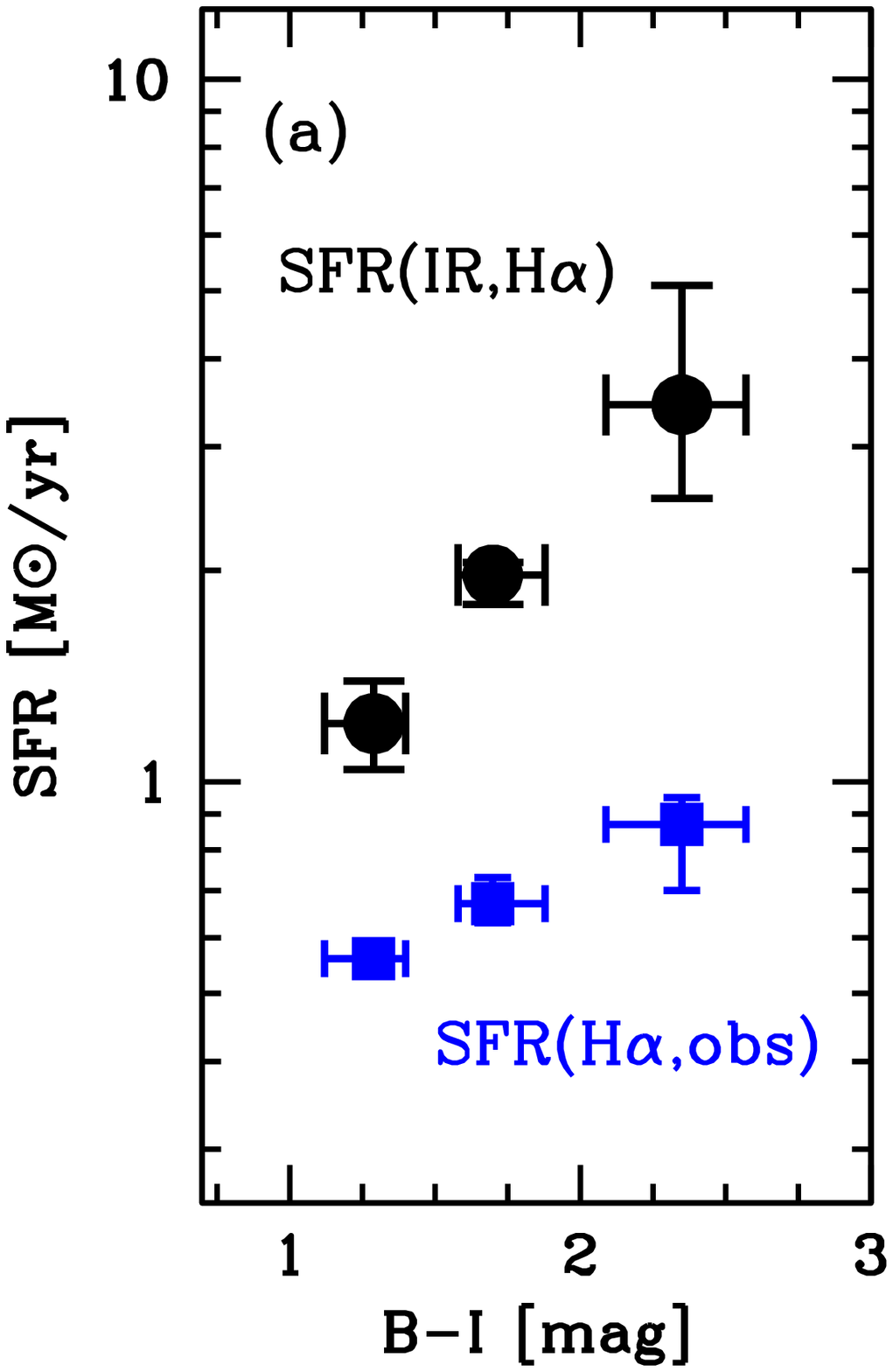}
    \hspace{-2.9cm}
    \epsfxsize 0.36\hsize
    \epsfbox{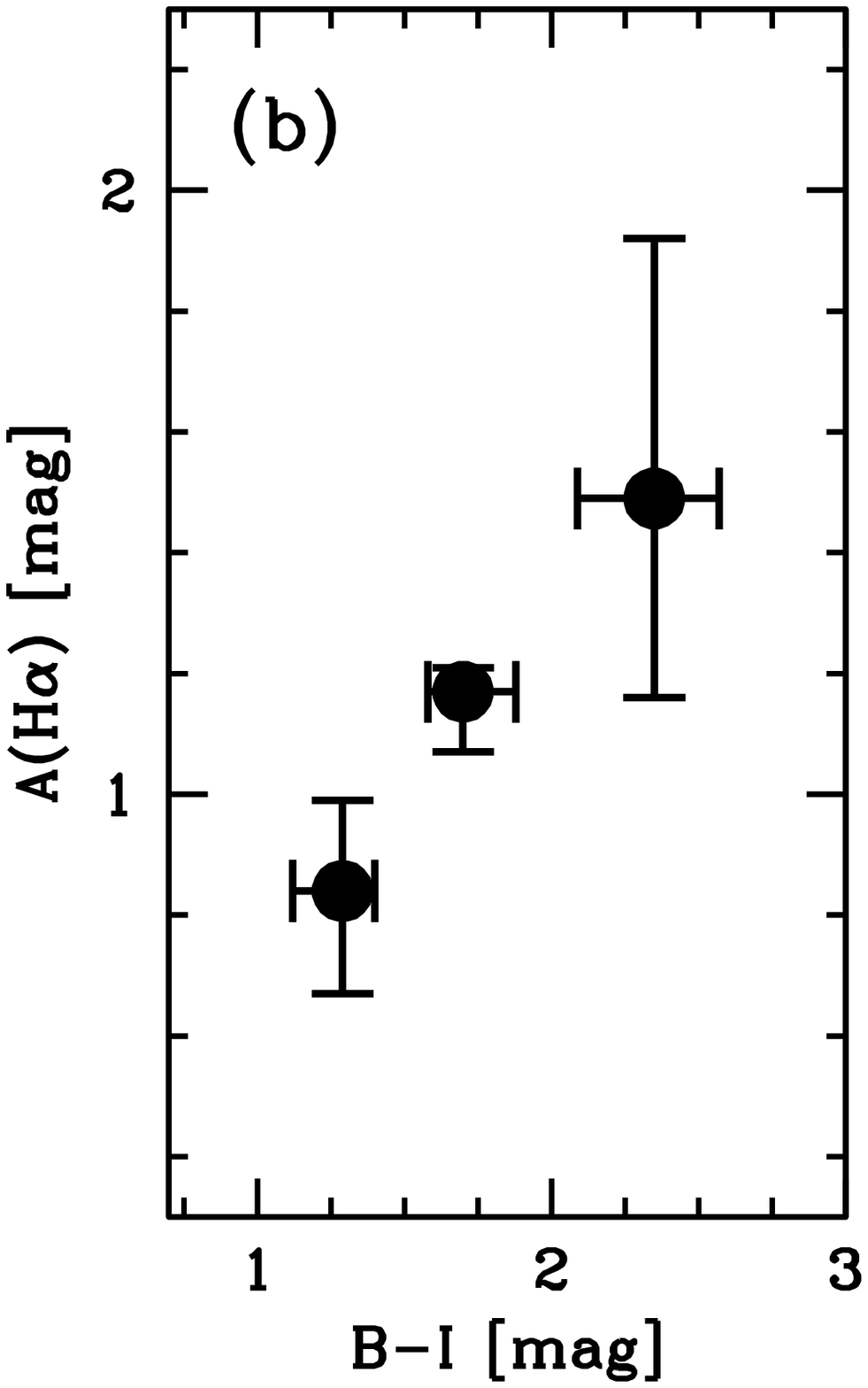}
    \hspace{-2.6cm}
    \epsfxsize 0.36\hsize
    \epsfbox{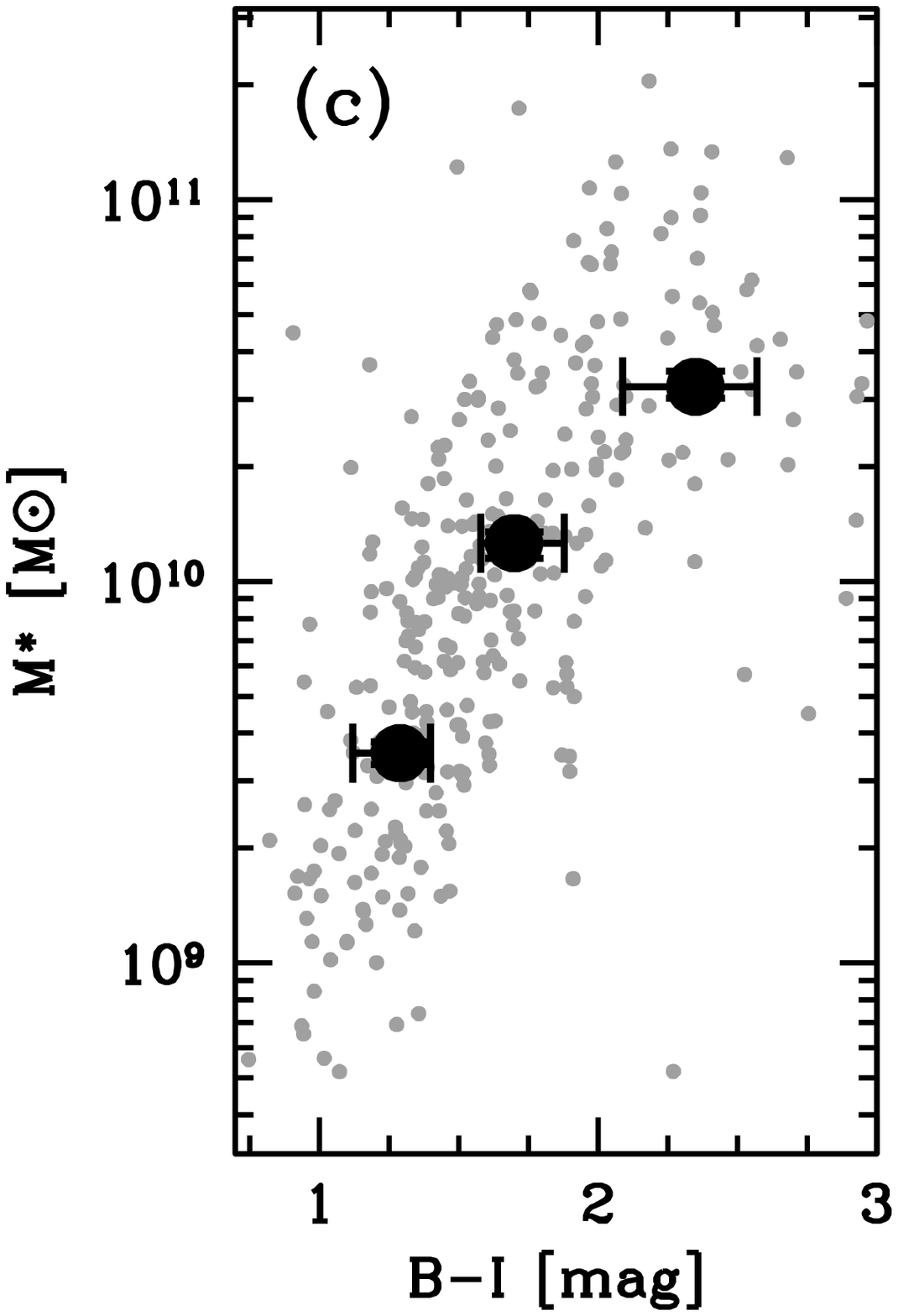}
    \hspace{-2.7cm}
    \epsfxsize 0.36\hsize
    \epsfbox{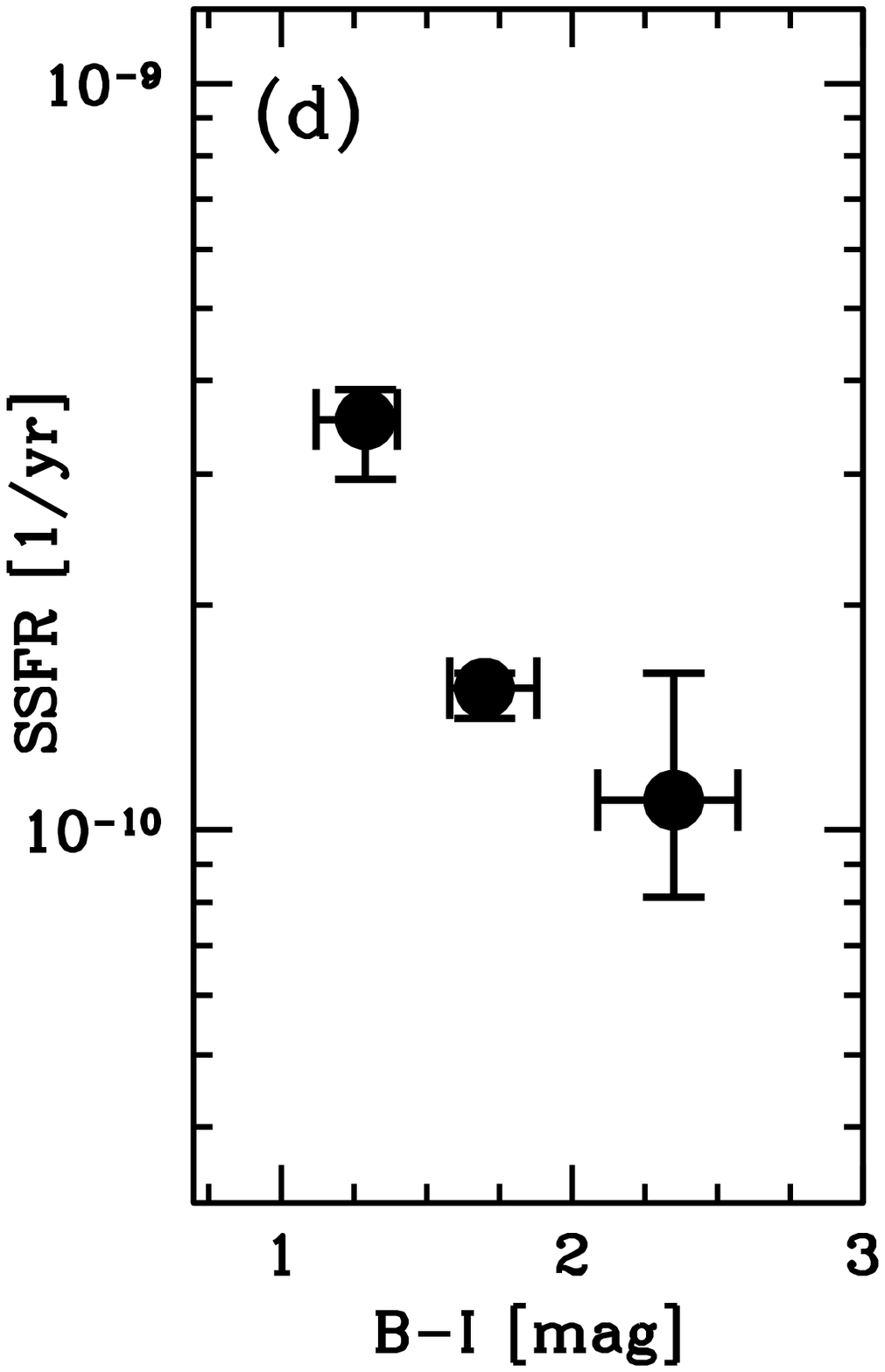}
   \end{center} 
  \vspace{-3mm}
   \caption{ The stacked properties of the H$\alpha$ emitters as a function of {\it B$-$I} colours (rest-frame {\it U$-$V}). The vertical error-bars show the 1$\sigma$ uncertainty derived from the bootstrap re-sampling approach during the MIPS stacking (see text), while the horizontal error-bars show the 25--75\% colour distribution of each subsample. (a) The SFR$_{\rm (IR,H\alpha)}$ derived from 24$\mu$m stacking analysis (black symbols). The blue symbols indicate the median values of SFR$_{\rm (H\alpha,obs)}$ (without dust extinction correction) for each subsample. (b) The $A_{{\rm H}\alpha}$ value for each subsample, calculated from the ratio of SFR$_{\rm (IR,H\alpha)}$ and SFR$_{\rm (H\alpha,obs)}$. (c) The median stellar mass for each colour subsample (black), as well as those for individual sources (grey dots). (d) The SSFR$_{\rm (IR,H\alpha)}$ for each subsample derived as SFR$_{\rm (IR,H\alpha)}$ normalized by stellar mass. These four plots demonstrate that the red H$\alpha$ emitters are massive star-forming galaxies with higher SFR and higher dust extinction compared with normal blue H$\alpha$ emitters. }
\label{fig:stack_color}
 \end{figure*}

\section{Results and Discussion}

\subsection{Panoramic H$\alpha$ and MIR view of the Cl\,0939 cluster}

In Fig.~\ref{fig:map_and_cmd}, we show the spatial distribution of the MIR-detected H$\alpha$ emitters and the MIR-detected photo-$z$ member galaxies. We also show all the H$\alpha$ emitters and photo-$z$ member galaxies, and the locations of the West Clump and the four surrounding groups following the definitions in \cite{koy11}. Note that we only show the galaxies located within the overlapped regions between our Subaru and MIPS FoVs. In the right-hand panel of Fig.~\ref{fig:map_and_cmd}, we show the colour--magnitude diagram for cluster (1~Mpc from the cluster centre), group (including the west clump and four groups), and field environments. These colour--magnitude diagrams show an overall trend that the MIPS-detected sources tend to be luminous ($z'$~$\lsim$~21~mag) and to have red optical colours ({\it B$-$I}~$\gsim$~2~mag).

We compare the {\it B$-$I} colour distribution of the H$\alpha$ emitters in each environment (see histograms in Fig.~\ref{fig:map_and_cmd}). The lack of the red H$\alpha$ emitters ({\it B$-$I}$>$2 mag) is clearly visible in the cluster environment, while a large number of H$\alpha$ emitters tend to have red colours in the group environment. The Kolmogorov-Smirnov (KS) test shows that the probability that the colour distributions of H$\alpha$ emitters in clusters and groups environment are from the same parent population is 1\%. The trend becomes less significant if we compare the group galaxies with field galaxies, but the KS test still suggests that the group and field galaxies are unlikely to be drawn from the same parent population (2.5\%). This probability goes down to $<$2\% if we use ``all'' H$\alpha$ emitters sample within the Suprime-Cam FoV (including the galaxies located outside the MIPS data coverage). Furthermore, we calculate the fraction of the red galaxies with any indication of star formation activity (i.e.\ H$\alpha$ or MIR detection): $f$$=$$N_{\rm  red{} SF}/N_{\rm all red}$. We find fractions of 4$\pm$1\%, 17$\pm$3\%, 11$\pm$1\% for cluster, group, and field environments, respectively. We therefore conclude that the red star-forming galaxies are most frequently seen in the group-scale environment (supporting our previous finding in \citealt{koy11}); in other words, a non-negligible fraction of optically red galaxies in the group environment are still actively forming stars.

  \begin{figure*}
   \vspace{-0.8cm}
   \begin{center}
    \leavevmode
    \epsfxsize 0.49\hsize
    \epsfbox{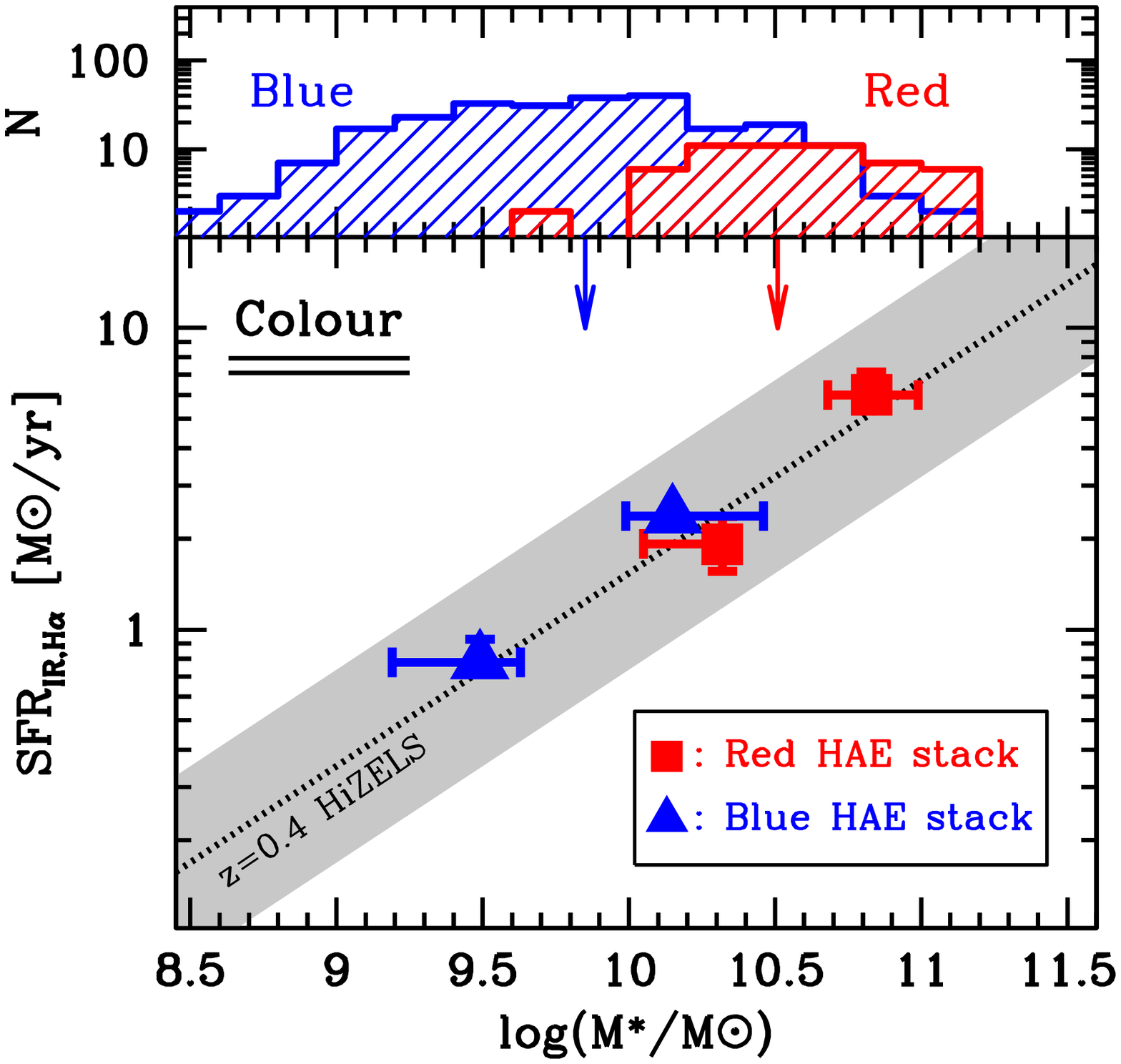}
    \hspace{-2mm}
    \epsfxsize 0.49\hsize
    \epsfbox{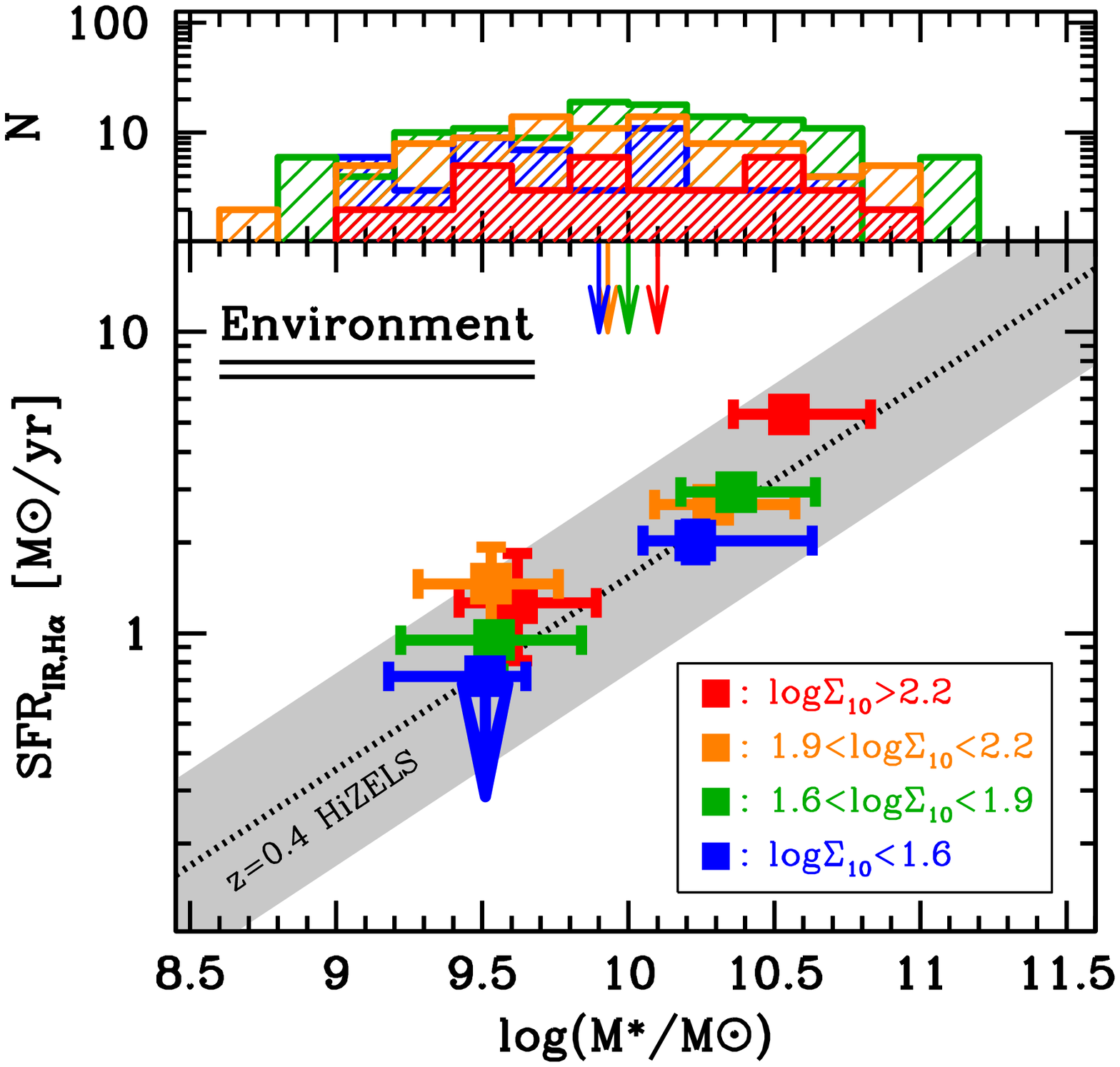}
   \end{center} 
  \vspace{-5mm}
   \caption{({\it Left}): The median SFRs from 24$\mu$m stacking analysis for the red and blue H$\alpha$ emitters as a function of stellar mass. The red and blue samples are further divided into two equal-sized stellar mass bins. The dotted line and the shaded region show the best-fitted SFR--$M_*$ relation and its scatter for the HiZELS $z=0.4$ H$\alpha$ emitter sample from Sobral et al. (2013), derived from SFR$_{\rm {(H\alpha,M_*)}}$. All our data points show an excellent agreement with the HiZELS relation. ({\it Right}): The same plot as the left-hand panel, but for our environmental subsamples (each sample is further divided into two equal-sized stellar-mass bins). In both panels, we show the $M_*$ distribution (histograms) and the median stellar mass (arrow) for each sample. In each plot, the vertical error-bars are from the bootstrap re-sampling in the MIPS stacking analysis, while the horizontal error-bars show the 25--75\% $M_*$ distribution for each subsample. }
\label{fig:ms_z04}
 \end{figure*}

\subsection{Stacked MIR properties as a function of galaxy colours}

The colour--magnitude diagrams in Fig.~\ref{fig:map_and_cmd} show that a fraction of the red H$\alpha$ emitters are individually detected at 24$\mu$m, suggesting they are dusty sources rather than passive galaxies. However, the limited depth of the 24$\mu$m data prevents us from assessing the general MIR properties of the faint H$\alpha$ galaxies. We therefore apply a 24$\mu$m stacking analysis to study the properties of the H$\alpha$ emitters more generally. To do this, we divide the full H$\alpha$ emitter sample in the Cl\,0939 field into three colour bins (at {\it B$-$I}$=$1.5 and 2.0), and perform stacking analysis as described in \S~2.1.3. The results are shown in Fig.~\ref{fig:stack_color}a. It is clear that the redder H$\alpha$ sources have higher SFRs than bluer sources. We also show in Fig.~\ref{fig:stack_color}a the SFRs derived from the median H$\alpha$ flux for each subsample, without dust extinction correction (SFR$_{\rm (H\alpha,obs)}$ hereafter). The same trend is still visible, but it is weaker than that obtained from MIPS stacking analysis. This result suggests that the redder H$\alpha$ galaxies are ``dustier'' than bluer galaxies. In Fig.~\ref{fig:stack_color}b, we show the extinction at H$\alpha$ derived from a ratio of the SFR$_{\rm (IR,H\alpha)}$ (from MIPS stacking) and the dust-free H$\alpha$ SFR (SFR$_{\rm (H\alpha,obs)}$) as a function of galaxy colours. This plot clearly shows that the red H$\alpha$ emitters have much higher extinction (with $A_{\rm{H}\alpha}$$\sim$1.5~mag) compared with the blue emitters ($A_{\rm{H}\alpha}$$\lsim$1 mag), suggesting that the red H$\alpha$ emitters are dusty, star-forming galaxies.

We should note that the clear trend that the redder H$\alpha$ emitters have higher SFRs could be produced by a stellar mass difference between red and blue H$\alpha$ emitters. In Fig.~\ref{fig:stack_color}c, we show the estimated $M_*$ for the individual H$\alpha$ emitters (see \S~2.3), as well as the median stellar mass for each colour subsample. The red H$\alpha$ emitters tend to have much higher stellar masses (by a factor of $\sim$10) than the blue H$\alpha$ emitters, which is more significant than the difference in SFR (i.e.\ Fig.~\ref{fig:stack_color}a). Consequently, it turns out that the red H$\alpha$ emitters tend to have lower specific star formation rate (SSFR) than blue H$\alpha$ galaxies (see Fig.~\ref{fig:stack_color}d), but the SSFRs for the red H$\alpha$ emitters are still at the $\sim$10$^{-10}$~yr$^{-1}$ level, suggesting they are part of the ``star-forming'' population.

As a further check, we show in Fig.~\ref{fig:ms_z04} (left) the results of our stacking results for the red and blue H$\alpha$ emitters on the SFR--$M_*$ diagram (we divide the red and blue H$\alpha$ emitter sample into two equal-sized stellar mass bins). Our data points show an excellent agreement with the best-fitted SFR--$M_*$ relation for the $z=0.4$ H$\alpha$ emitters from HiZELS (\citealt{sob13}), further supporting our conclusion that the red H$\alpha$ emitters are dusty star-forming galaxies, rather than dust-free passive galaxies. We recall that when deriving SFR--$M_*$ relation for the HiZELS sample, we adopt an $M_*$-dependent dust extinction correction to HiZELS data (see \S~2.3). The excellent agreement (over the wide $M_*$ range) between the SFR$_{\rm (IR,H\alpha)}$ and those from the independent, extinction corrected H$\alpha$ fluxes (SFR$_{\rm H\alpha,M_*}$) also supports the validity of our procedure for the dust extinction correction applied in this study.

\subsection{Environmental dependence} 

We have shown that our H$\alpha$ emitters tend to be located on the general SFR--$M_*$ sequence (Fig~\ref{fig:ms_z04}-left). An interesting question here is: does environment influence the SFR--$M_*$ relation? To answer this question, we attempt a similar analysis to \S~3.2 but dividing the whole $z=0.4$ H$\alpha$ emitter sample into four environment bins based on the local galaxy density (at $\log\Sigma_{10}=$1.6, 1.9, 2.2). The density is calculated using all cluster member galaxies (photo-$z$ selected and H$\alpha$ selected) with the nearest-neighbour approach, calculated within a radius to the 10\,th-nearest neighbour from each source. We further divide each environment subsample into two equal-sized stellar mass bins, and perform the 24$\mu$m stacking analysis. The results are shown in Fig.~\ref{fig:ms_z04} (right). The stacking analysis becomes challenging, particularly for the lower-mass sources due to the limited sample size, but broadly speaking, all our data points are likely to be located on the same SFR--$M_*$ sequence.  This suggests that the SFR--$M_*$ (or SSFR--$M_*$) relation for star-forming galaxies does not strongly correlate with the environment at $z=0.4$. We note that there seems to be a small ($\lsim$0.2~dex) positive offset for the highest-density bin on this SFR--$M_*$ diagram. Although this is not very significant, we will discuss this issue later in this subsection.

\begin{figure}
  \begin{center}
  \vspace{-1.5cm}
    \rotatebox{0}{\includegraphics[width=8.6cm,height=8.6cm]{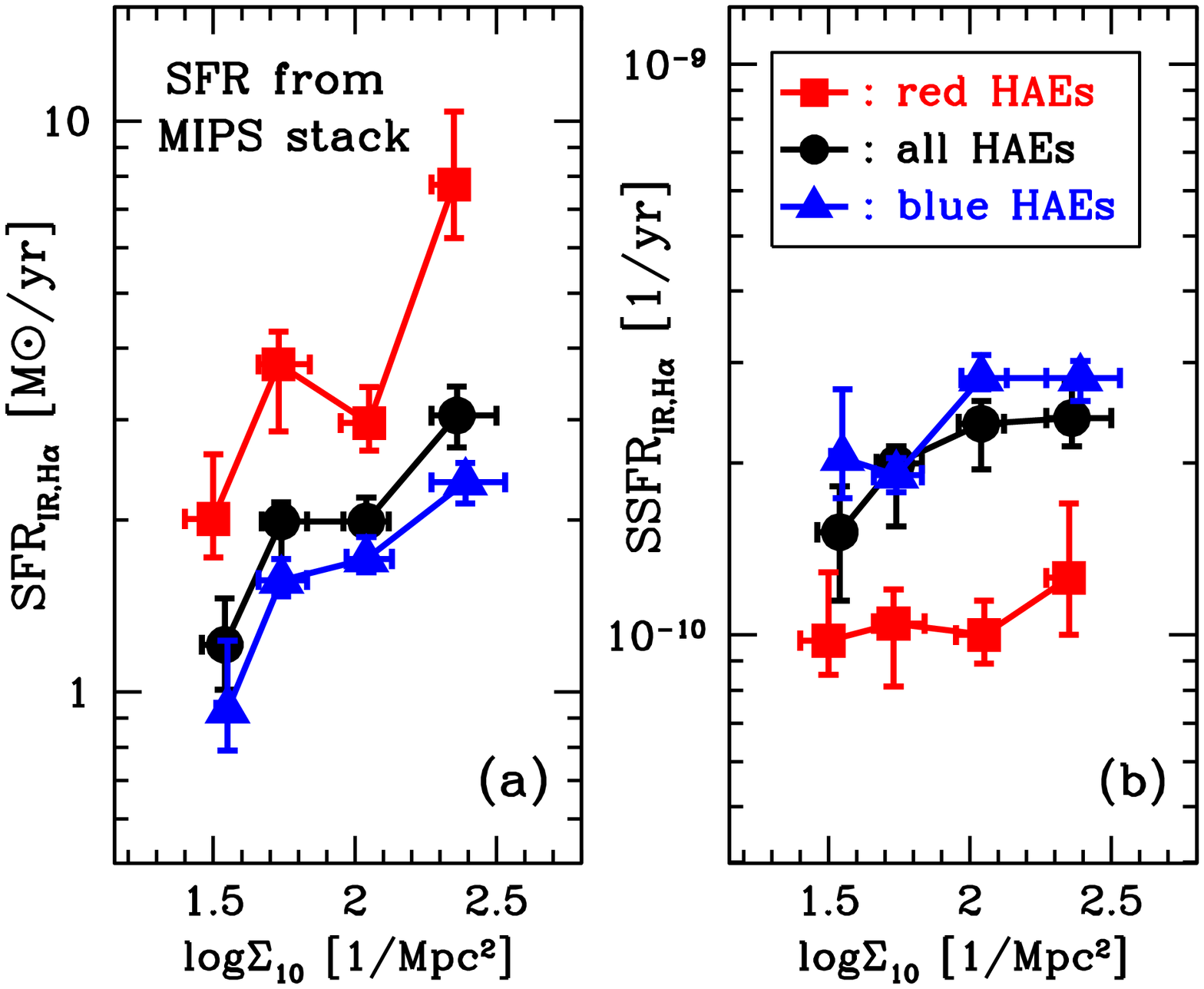}}
    \rotatebox{0}{\includegraphics[width=8.6cm,height=8.6cm]{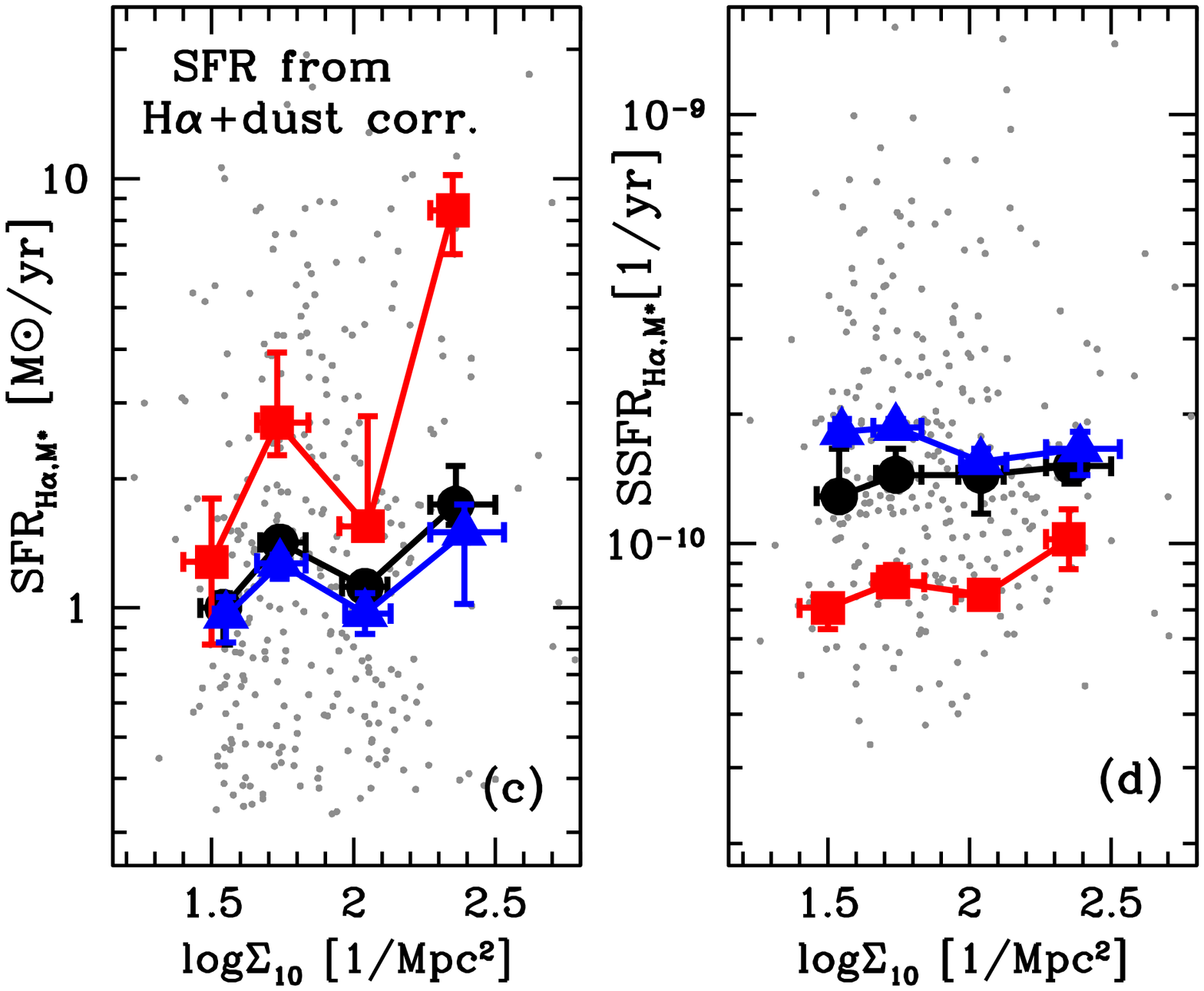}}
  \end{center}
  \vspace{-1.8cm}
  \caption{ The SFRs and SSFRs for our H$\alpha$ emitters as a function of the local galaxy density ($\Sigma_{10}$). The density is calculated with all member galaxies (photo-$z$ members and H$\alpha$ emitters), while we include only H$\alpha$ emitters in the SFR analysis presented here. The top panels (a and b) show the results from MIPS stacking, while the bottom panels (c and d) show those from H$\alpha$ with dust extinction correction. The line-connected circles, squares, and triangles with error-bars are the median SFR/SSFR for all, red, and blue galaxies in each environment bin, respectively. The vertical error-bars show the 1-$\sigma$ distribution from the bootstrap resampling for all cases, and the horizontal error-bars show the 25--75\% distribution for each subsample.
}
\label{fig:sfr_vs_env}
\end{figure}

Using the same environmental subsamples, we also test the SFR--density relation for star-forming galaxies at $z=0.4$. In Fig.~\ref{fig:sfr_vs_env}a, we show the SFR$_{\rm (IR,H\alpha)}$ (from 24$\mu$m stacking analysis) as a function of galaxy density. It is interesting to note that the SFR$_{\rm (IR,H\alpha)}$ {\it increases} with increasing galaxy density (by a factor of $\sim$3--4), and the trend is confirmed for both red and blue H$\alpha$ emitter samples. It should be noted that the star-forming activity of galaxies presented here only focuses on star-forming galaxies, and that we do not include passive galaxies in our analysis. Indeed, we showed in \cite{koy11} that the H$\alpha$ emitter ``fraction'' is a strong function of environment in this CL\,0939 field, showing a significant decline towards the cluster core. The readers should not be confused about this point --- the important message from our current analysis is that the SFRs of the uniformly H$\alpha$-selected galaxies at $z=0.4$ do show an environmental dependency in the sense that galaxies in high-density environment have higher SFRs. This enhancement of SFRs in high-density environment amongst star-forming galaxies can (at least partly) contribute to the ``reversal'' of the SFR--density relation in the distant Universe claimed by recent studies (e.g.\ \citealt{elb07}; \citealt{coo08}). 

In contrast to this, the trend becomes much less significant when we normalise the SFRs by $M_*$ to compare the specific SFR (SSFR; see Fig.~\ref{fig:sfr_vs_env}b). It is clear that the SSFR shows a weaker environmental trend than SFR, implying that the SFR excess detected in the high-density environment would largely be explained by the $M_*$ difference between the different environments. We note, however, that the $M_*$ distribution amongst H$\alpha$ emitters does not seem to be strongly dependent on environment (as shown by histograms in the right panel of Fig.~\ref{fig:ms_z04}). We can still find a weak trend that the galaxies in high-density environment tend to be more massive, but the difference is at the $\sim$0.2~dex level at maximum. The weakness or lack of the environmental dependence of the stellar mass distribution amongst star-forming galaxies is consistent with some recent studies (\citealt{gre12}; \citealt{gio12}), but this small $M_*$ difference may not be able to fully account for the significant SFR increase towards high-density environment. We therefore speculate that the SFR excess in the high-density regions could be explained by a ``mixed effect'' of both slightly higher stellar masses {\it and} a small SSFR excess (both at $\sim$0.2~dex level) in the high-density environment. Indeed, the small SSFR excess is visible as the remaining positive slope in the SSFR$_{\rm (IR,H\alpha)}$ versus $\log\Sigma_{10}$ plot (see Fig.~\ref{fig:sfr_vs_env}b), which is also equivalent to the small positive offset of the stacked data points for high-density environments in the SFR--$M_*$ plot (Fig.~\ref{fig:ms_z04} right). 

Interestingly, the increase of SSFR towards high-density environments is {\it not} visible when we use SFRs from H$\alpha$ alone (SFR$_{\rm (H\alpha,M_*)}$). In Fig.~\ref{fig:sfr_vs_env}cd, we show the same analysis using the SFR$_{\rm (H\alpha,M_*)}$. We can still see the increase of SFR towards high-density regions, but the trend for the SSFR becomes even flatter than that we derived from the SFR$_{\rm (IR,H\alpha)}$, showing an apparently contradicting result to Fig.~\ref{fig:sfr_vs_env}b. Although the difference is $\lsim$0.2~dex level, it would be interesting to investigate the origin of this different result more in detail. Naively, the different result from different SFR measurements may reflect a possible environmental variations of ``dustiness'' of galaxies. We remind that we apply the $M_*$-dependent extinction correction using the $A_{\rm H\alpha}$--$M_*$ relation established in the local Universe. As we showed in \S~2.3, the SFR$_{\rm (H\alpha,M_*)}$ agrees well with the SFR$_{\rm (IR,H\alpha)}$ (at least in an average sense), but it might be too simplistic to assume this $M_*$-dependent extinction correction to galaxies residing in all environments. 

  \begin{figure*}
   \vspace{-1.3cm}
   \begin{center}
    \leavevmode
    \hspace{5mm}
    \epsfxsize 0.50\hsize
    \epsfbox{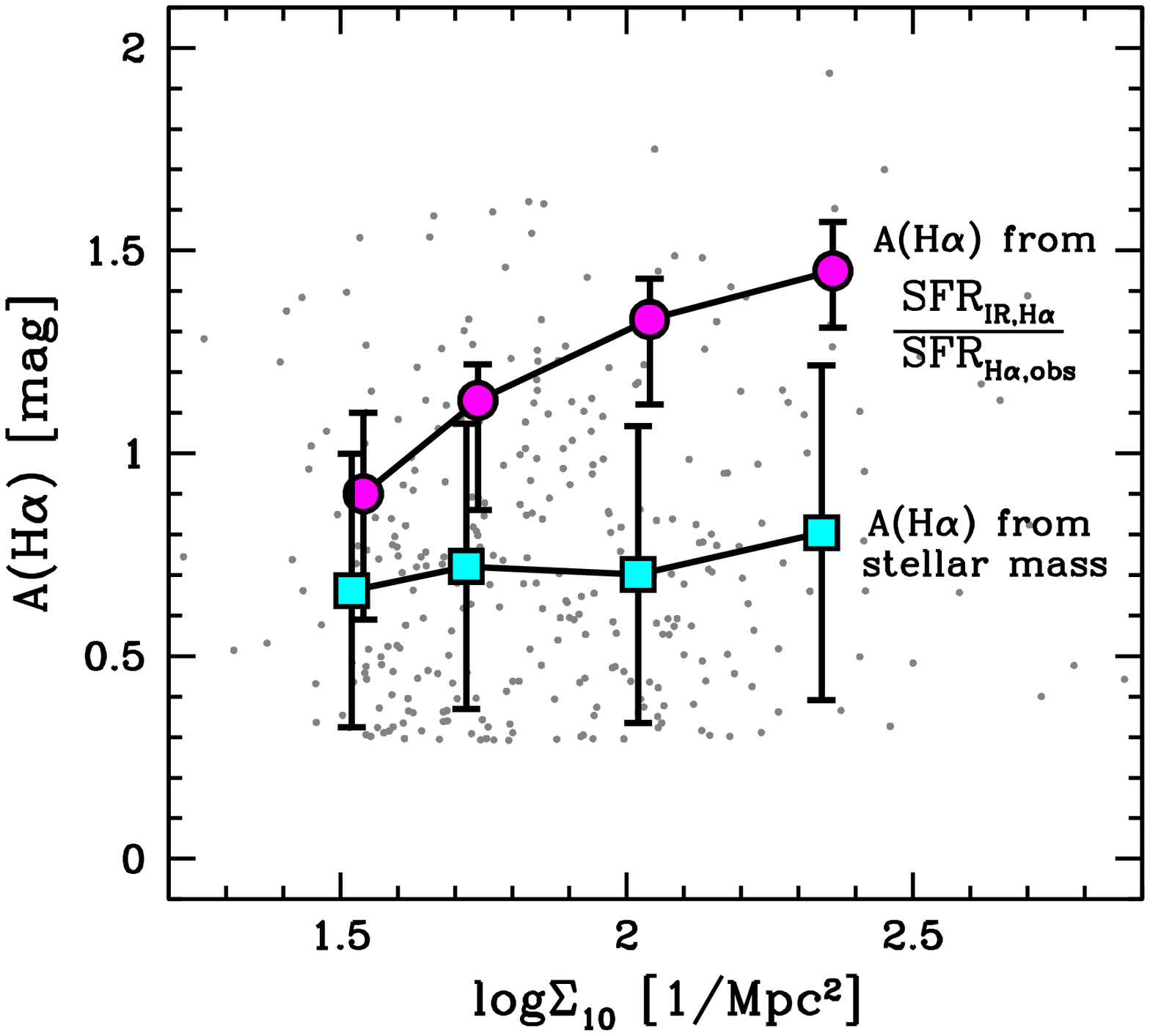}
    \hspace{-7mm}
    \epsfxsize 0.50\hsize
    \epsfbox{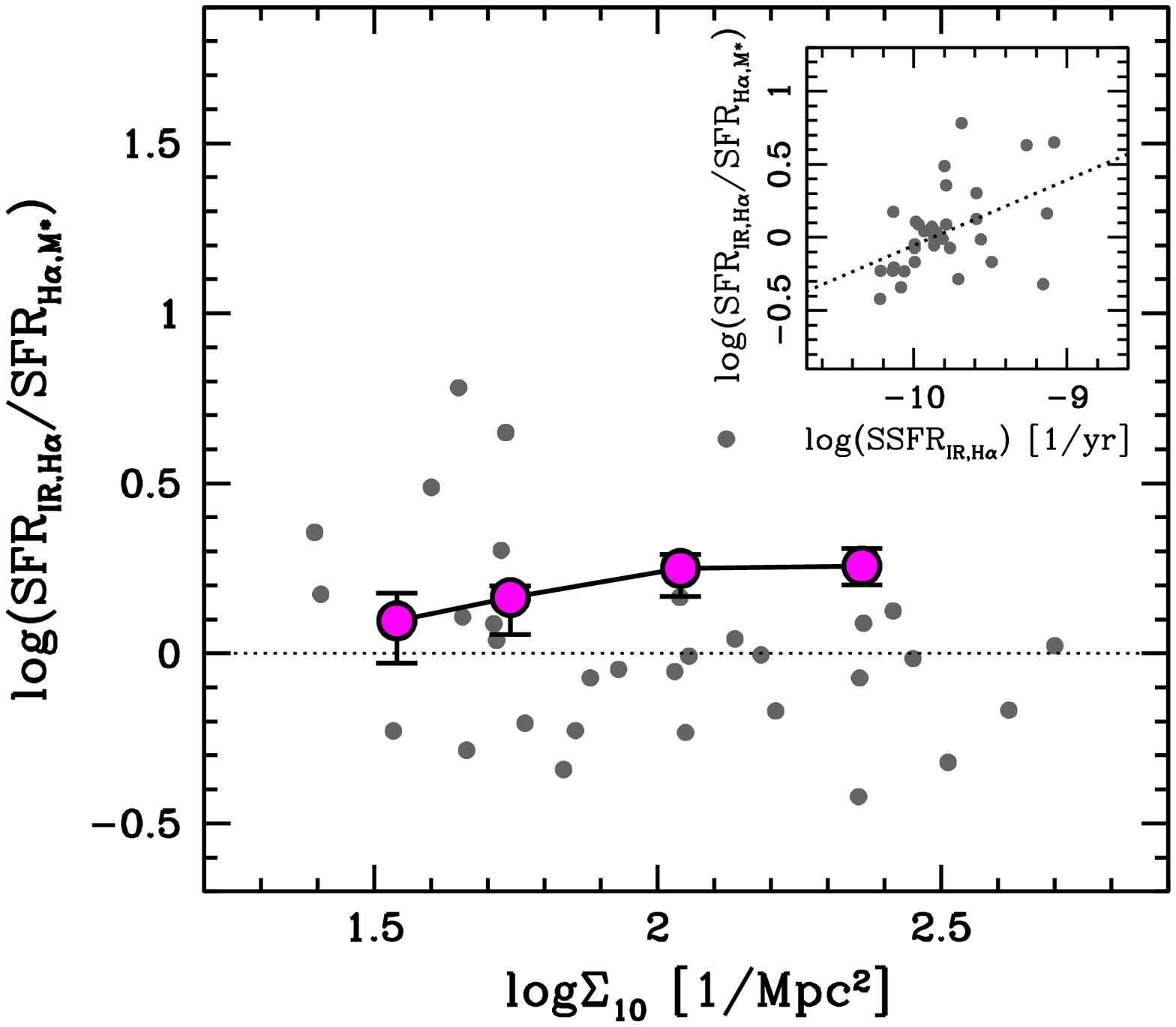}
   \end{center} 
  \vspace{-7mm}
   \caption{({\it Left}): The dust extinction ($A_{\rm H\alpha}$) of our $z=0.4$ H$\alpha$ emitters (in the Cl\,0939 field) as a function of environment. The line-connected circles show the estimate from MIPS stacking analysis (via SFR$_{\rm (IR,H\alpha)}$/SFR$_{\rm (H\alpha,obs)}$), and the line-connected squares show the median of $A_{\rm H\alpha}$ derived from $M_*$ in each density bin. The error-bars show the 1-$\sigma$ distribution for each subsample. The grey dots show $A_{\rm H\alpha}$ of individual H$\alpha$ emitters (derived from $M_*$). ({\it Right}): A similar plot to the left-hand panel, but showing the ratio between the SFR$_{\rm (IR,H\alpha)}$ and the SFR$_{\rm (H\alpha,M_*)}$ as a function of environment. The grey points show MIPS-detected H$\alpha$ emitters, while the line-connected circles show the results from MIPS stacking for each environment subsample. In the inset, we show the ratio of SFR$_{\rm (IR,H\alpha)}$/SFR$_{\rm (H\alpha,M_*)}$ as a function of SSFR$_{\rm (IR,H\alpha)}$ for the MIPS-detected H$\alpha$ emitters, showing a weak trend that higher SSFR galaxies tends to have higher SFR$_{\rm (IR,H\alpha)}$/SFR$_{\rm (H\alpha,M_*)}$ ratio. The dotted line shows the best-fitted relation for the plotted data points. These two plots demonstrate that the star-forming galaxies in high-density environment tend to be dustier, and the SFR$_{\rm (H\alpha,M_*)}$ could be underestimated in such extreme environments (by up to $\sim$0.2~dex.) 
}
\label{fig:AHa_vs_density}
 \end{figure*}

To test this possibility, we show in Fig.~\ref{fig:AHa_vs_density} (left) the median $A_{\rm H\alpha}$ value in each environmental subsample. We estimate $A_{\rm H\alpha}$ with two independent methods; (1) from SFR$_{\rm (IR,H\alpha)}$/SFR$_{\rm (H\alpha,obs)}$ and (2) from stellar mass. Assuming that the SFR$_{\rm (IR,H\alpha)}$ can provide more reliable measurements, Fig.~\ref{fig:AHa_vs_density} (left) suggests that the star-forming galaxies in high-density environments tend to be dustier (by $\sim$0.5~mag at maximum), whereas this trend is not visible for the $A_{\rm H\alpha}$ derived from $M_*$. Similarly, we show in Fig.~\ref{fig:AHa_vs_density} (right) a more direct comparison between SFR$_{\rm (IR,H\alpha)}$ and SFR$_{\rm (H\alpha,M_*)}$. These two SFRs are consistent within error-bars in the low-density environments ($\log\Sigma_{10}$$\lsim$2.0 where most of the galaxies reside), but we tend to underestimate SFR$_{\rm (H\alpha,M_*)}$ for galaxies in high-density environment at the 0.1--0.2~dex level. This would be the right answer to why the two different SFR indicators provide apparently different results in Fig.~\ref{fig:sfr_vs_env}. It may be possible that the increasing dust extinction with the increasing galaxy number density could be (at least partially) driven by different timescales of the SFR indicators; H$\alpha$ is more sensitive to the shorter time scale of star formation than MIR. However, the SFR$_{\rm (IR,H\alpha)}$ derived by the combined IR+H$\alpha$ approach is reported to show a tight correlation between the SFRs derived from H$\alpha$ with extinction correction based on the H$\beta$/H$\alpha$ ratio (\citealt{ken09}), so we expect this effect should be small.

The physical interpretation of the above result may be straightforward. In high-density environments, such as clusters or groups, galaxy--galaxy interactions/mergers or gas/dust stripping should happen more frequently, and these environmental effects probably result in a more compact (and more obscured) configuration of the star formation taking place within those galaxies (hence exhibiting higher SSFR). We test this hypothesis by plotting SFR$_{\rm (IR,H\alpha)}$/SFR$_{\rm (H\alpha,M_*)}$ ratio as a function of SSFR$_{\rm (IR,H\alpha)}$ (see the inset in Fig.~\ref{fig:AHa_vs_density} right). Although the number of H$\alpha$ emitters individually detected at 24$\mu$m is small, there seems to be a weak trend that galaxies with higher SSFR tend to have higher SFR$_{\rm (IR,H\alpha)}$/SFR$_{\rm (H\alpha,M_*)}$ ratio. We note that this result seems to be qualitatively consistent with some recent studies. For example, \cite{sob11} find that a higher fraction of star-forming galaxies are associated with mergers in higher-density environments, using their $z=0.8$ H$\alpha$ emitters sample from HiZELS. Also, \cite{gea09} performed MIR spectroscopy of luminous infrared galaxies (LIRGs) in $z=0.4$ cluster environments using IRS/{\it Spitzer}, and find that their MIR SEDs more resemble nucleated (dusty) starbursts (rather than star-forming disc). They propose these dusty starbursts in distant cluster environments are the progenitors of the bulge-rich, local cluster S0s. \cite{raw12} use {\it Herschel} data to find "warm-dust" galaxies in $z\sim 0.3$ cluster environment. \cite{raw12} propose that these galaxies will be explained by a "dust-stripping" mechanism by cluster environments; i.e.\ cool dust in the outskirts of galaxies are more easily stripped, resulting in the warm-dust population in cluster environments. Although the number of studies which focus on the dust properties of cluster galaxies is currently very limited, these studies would support our finding that the galaxies in high-density environment tend to be dustier. We therefore speculate that the preference of such dusty galaxies in high-density environments create the trend of increasing SSFR$_{\rm (IR,H\alpha)}$ towards high-density environments (as we showed in Fig.~6b), whilst keeping the SSFR$_{{\rm (H\alpha},M_*)}$ versus $\Sigma_{10}$ relation flat (as shown in Fig.~6d). Based on the possible environmental variations of $A_{\rm H\alpha}$, we incorporate this ``environmental uncertainty'' in the dust extinction correction ($\sim$0.5~mag) in the following discussions.

 \begin{figure*}
  \begin{center}
    \vspace{-5.3cm}
    \leavevmode
    \rotatebox{0}{\includegraphics[width=16.5cm,height=16.5cm]{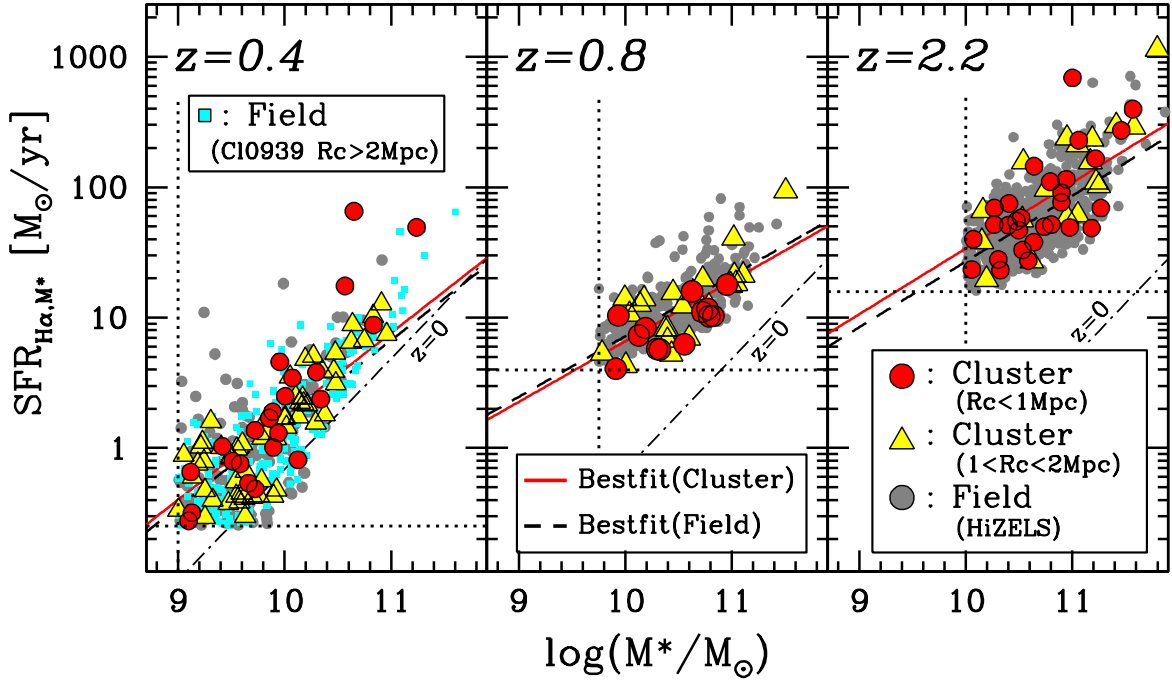}}
  \end{center}
  \vspace{-2.5cm}
  \caption{ The SFR--$M_*$ relation for the H$\alpha$-selected galaxies at $z=0.4$, $0.8$, $2.2$. In each panel, we plot the H$\alpha$ emitters in cluster environment located within $R_c<$1~Mpc and at 1$<R_c<$2~Mpc from the cluster centre. We also plot the H$\alpha$ emitters in general field environment selected from HiZELS at the same redshifts. Note that all the H$\alpha$ emitters plotted here are selected as those having EW(H$\alpha$+[{\sc Nii}])$_{\rm rest}$$>$30\AA. The dotted lines show the SFR and $M_*$ cut applied at each redshift. The dashed line shows the best-fitted SFR--$M_*$ relation for the HiZELS sample, while the solid line shows the relation for the cluster sample assuming the same slope as the HiZELS relation. The dot--dashed line is the local ($z=0$) relation derived from the equation provided by Whitaker et al. (2012). It is clear that the SFR--$M_*$ relation evolves significantly since $z\sim2$ in both cluster and field environment, while at fixed redshifts, the environmental dependence of the SFR--$M_*$ relation seems to be very small, at least when we consider the SFRs derived from H$\alpha$ emissions. }
  \label{fig:ms_evolution}
 \end{figure*}

\subsection{The evolving SFR--M$_*$ relation since z$\sim$2}

We have examined the star-forming activity of galaxies and its dependence on stellar mass and environment at $z=0.4$. Here, we discuss the environmental dependence of the evolution of star-forming galaxies in a broader context, particularly focusing on the evolution and environmental dependence of the SFR versus $M_*$ relation across cosmic time. By compiling all of the H$\alpha$ emitter samples together (including our MAHALO and HiZELS samples; see \S~2.2), we examine the SFR versus $M_*$ sequence in cluster and field environments at $z=0.4$, $0.8$, and $2.2$. We note that, while all the samples are selected based on the H$\alpha$ line, the EW cut applied in each survey is slightly different. Therefore we decide to apply the same (rest-frame) EW cut to all the samples, EW$_{\rm  rest}$(H$\alpha$+[{\sc Nii}])$=$30\AA, down to which all our H$\alpha$ data are complete. We also note that it is not possible to make a strictly fair comparison between clusters at different redshifts; for example, a density-based definition of environment requires accurate membership determination, while using the virial radius could be misleading because our high-$z$ clusters are not virialised yet. Therefore, we instead use galaxies within 2~Mpc (in physical scale) from each cluster centre as ``cluster'' galaxies in the following discussion. We note again that in the remaining of this paper we use the H$\alpha$-based SFRs (SFR$_{\rm H\alpha,M_*}$), which could be environmentally uncertain at the $\sim$0.5 mag level (as we showed in the previous subsection). However, it is important to investigate the presence (or lack) of any environmental variations in the SFR--$M_*$ relation across cosmic time, based on our largest H$\alpha$ emitter samples ever available.

In Fig.~\ref{fig:ms_evolution}, we show all the H$\alpha$-selected galaxies in clusters and field environments at each redshift. We use different symbols for the cluster H$\alpha$ emitters located within $R_c<$1~Mpc, 1$<$$R_c$$<$2~Mpc, and the field H$\alpha$ emitters from HiZELS (see labels in the plot). For $z=0.4$, we also show the H$\alpha$ emitters located at $R_c>$2~Mpc from the Cl\,0939 cluster. This plot clearly shows that the SFR--$M_*$ relation evolves with redshift, while the relation is always independent of environment out to $z\sim 2$, qualitatively consistent with the situation in the local Universe (e.g. \citealt{pen10}; \citealt{wij12}). We also show the best-fitted SFR--$M_*$ relation in each panel. It can be seen that the offset between the relation for cluster and field galaxies is always small. The slope of the SFR--$M_*$ relation tends to be steeper for the lower-redshift samples, which is also qualitatively consistent with previous works (e.g.\ \citealt{whi12}), but we note that the best-fit relation drawn on the plot is uncertain because our star-forming galaxy samples are not completely stellar-mass limited.

\begin{figure}
  \begin{center}
    \leavevmode
    \vspace{-0.5cm}
    \rotatebox{0}{\includegraphics[width=8.4cm,height=8.4cm]{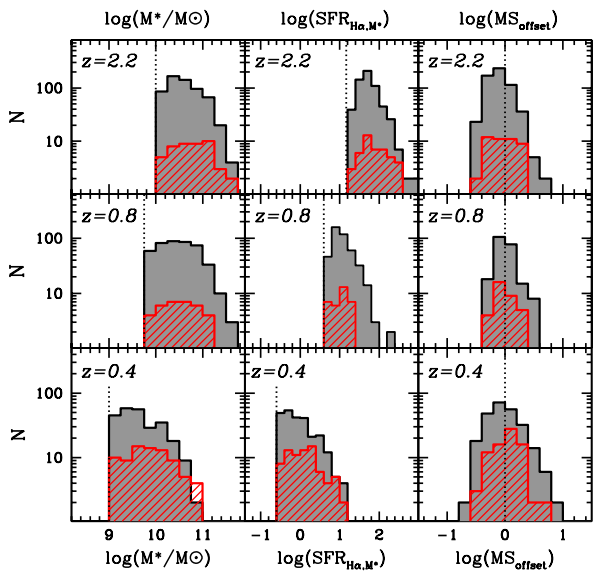}}
  \end{center}
  \vspace{-0.5cm}
  \caption{ The distribution of $M_*$ (left), SFR (middle), and offsets from the main sequence of field galaxies (right) at each redshift. The shaded histograms show the results for HiZELS sample, while the hatched histograms show the results for cluster ($R_c<2$~Mpc) galaxies. The vertical dotted lines in the left and middle panels show the $M_*$ or SFR cut we applied for each redshift sample, while the dotted lines in the right-hand panels show the location of the zero-offset. The actual difference between cluster and field galaxies is always small ($\lsim$0.1--0.2~dex at maximum), but we note that a statistical test suggests that the two distributions may be different for our $z=2.2$ sample in the sense that the cluster galaxies have higher $M_*$ and higher SFR (see text). }
  \label{fig:histograms}
\end{figure}

We here examine the galaxy distribution on the SFR--$M_*$ plane more in detail. In Fig.~\ref{fig:histograms}, we show the distributions of $M_*$, SFR, and the offset from the main sequence (best-fitted SFR--$M_*$ for field galaxies). The shaded histograms are for field H$\alpha$ emitters, and the hatched histograms are for cluster H$\alpha$ emitters. Broadly speaking, it seems that there is no significant environmental difference at any of the three epochs. For $z=0.4/0.8$, the KS test suggests that it is unlikely that the cluster and field galaxies are from a different parent population, while for $z=2.2$, we find a possible trend that cluster galaxies have a small excess in all three properties (the KS test actually shows $<$1\% probability that the cluster and field samples are from the same parent population). We note that this trend is qualitatively consistent with some earlier studies showing a higher stellar masses in star-forming galaxies in $z>2$ proto-clusters (e.g. \citealt{ste05}; \citealt{kur09}; \citealt{hat11}; \citealt{mat11}; \citealt{koy13}). Our current analysis also supports the idea that the star-forming galaxies in proto-cluster environment tend to be more massive than the general field galaxies, and this $M_*$ excess would also account for the SFR excess in the proto-cluster environment; this may represent an accelerated galaxy growth in the early phase of the cluster assembly history.

\begin{figure}
  \begin{center}
    \leavevmode
    \vspace{-0.6cm}
    \rotatebox{0}{\includegraphics[width=8.9cm,height=8.9cm]{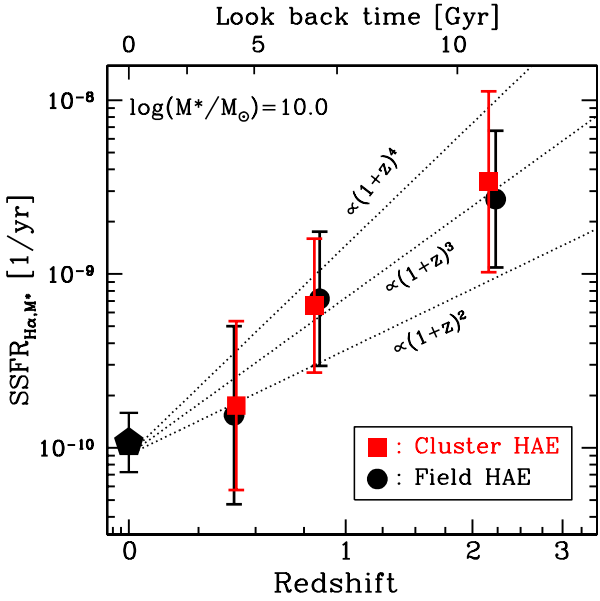}}
  \end{center}
  \vspace{-0.8cm}
  \caption{ The redshift evolution of the (H$\alpha$-derived) SSFR at $M_*=10^{10}M_{\odot}$ derived from the best-fitted SFR--$M_*$ relation for cluster (red squares) and field (black circles) galaxies. The error-bars incorporate the standard deviation around their best-fit SFR--$M_*$relation (see Fig.~8), and the maximum environmental uncertainty in $A_{\rm H\alpha}$ (0.5~mag; see Fig.~7). The dotted lines are the evolutionary tracks following $\propto (1+z)^2$, $\propto (1+z)^3$, and $\propto (1+z)^4$, to guide the eye. The local data point is derived by adopting $z=0$ in the equation of Whitaker et al. (2012).}
  \label{fig:ssfr_evolution}
\end{figure}

Finally, we quantify the evolution of the star-forming activity of star-forming galaxies in the cluster and field environment. In Fig.~\ref{fig:ssfr_evolution}, we plot the redshift evolution of the SSFR$_{\rm (H\alpha,M_*)}$ of H$\alpha$ emitters in clusters (red squares) and in field environment (black circles) at the stellar mass of $\log(M_*/M_{\odot})=$10. It is found that the SSFR$_{\rm (H\alpha,M_*)}$ of H$\alpha$-selected galaxies evolves significantly, going approximately as $(1+z)^3$, since $z\sim 2$ in both clusters and the field. This decline of SSFR is in good agreement with many studies of the cosmic star formation history (e.g.\ \citealt{yos06}; \citealt{kar11}; \citealt{sob13}). Therefore, an important indication from this study is that the evolution of star-forming galaxies in cluster environments seems to be following the same evolutionary track as that of general field galaxies, as far as we consider the SFRs derived from H$\alpha$ emissions.

\subsection{Comparison with other studies}

Studying the environmental dependence of galaxy star formation activity in the distant Universe is obviously an important step towards understanding the physical processes which drive the environmental effects. Since discussion on the ``reversal'' of the SFR--density relation was invoked by \cite{elb07}, there has been much debate about the role of environment in the distant Universe. In this paper, we reported that the SFR tends to be higher in higher-density environments at $z=0.4$ (\S~3.3), and we expect that this enhancement of SFRs amongst star-forming galaxies in high-density environment is at least partially responsible for the reversal of the SFR--density relation in the distant Universe. We note that our results are qualitatively consistent with some recent studies. For example, \cite{sob11} used H$\alpha$ emitters sample at $z=0.8$ selected from HiZELS to show higher median SFRs (by a factor of $\sim$2--3) in high-density environment compared to low-density environment. They also showed that the stellar mass of H$\alpha$ emitters is weakly correlated with the environment (with $\sim$0.3~dex increase in their highest-density bins), which also agrees with our finding in \S~3.3. \cite{tra09} analysed MIR data of a ``super-group'' environment at $z=0.37$ to show that the characteristic IR luminosity ($L^*$) in the group environment is higher than that in the field, based on the analysis of the IR luminosity function (see also \citealt{chu10}). Related to this, some studies of distant clusters show a peak of star formation activity at a certain galaxy density which corresponds to group or cluster outskirts environment (e.g.\ \citealt{pog08}; \citealt{koy10}; \citealt{gea11}).

On the other hand, we find that the environmental dependence of the SFR--$M_*$ relation is always small since $z\sim 2$ ($\lsim$0.2~dex at maximum), even if we take the possible environmental uncertainty in the dust extinction correction into account. In fact, a growing number of studies recently have reported a weakness or absence of any relation between SSFR and environmental density amongst star-forming galaxies at least out to $z\sim 1$, or possibly to $z\sim 2$. In the local Universe, \cite{bal04} showed that the EW(H$\alpha$) distribution (equivalent to SSFR distribution) amongst star-forming galaxies is independent of environment. More recent studies also indicated that the SFR--$M_*$ relation for local star-forming galaxies does not correlate with the environment (e.g.\ \citealt{pen10}; \citealt{wij12}). Similar suggestions have also been made for distant star-forming galaxies as well. For example, \cite{mcg11} studied a large sample of $z=0.4$ group galaxies to show that the average SSFRs of star-forming galaxies are the same in groups as in field environments. \cite{muz12} also showed that SSFR of star-forming galaxies is independent of environment at fixed stellar mass from their detailed spectroscopic survey of $z\sim 1$ cluster galaxies (see also \citealt{gre12}). Furthermore, our recent studies of distant (proto-)clusters also find a hint that the SSFR of star-forming galaxies is independent of environment at fixed stellar mass out to $z\sim 2.5$ (\citealt{tad12}; \citealt{hay11}; 2012; \citealt{koy13}). 

However, it should be noted that the independence of the SFR--$M_*$ relation for star-forming galaxies with environment in the distant Universe is still controversial (\citealt{pat11}; \citealt{vul10}; \citealt{li11}). Indeed, this kind of analysis could be highly sensitive to the sample selection, the measurement of star formation rates, and the definitions of environment (as we showed in \S~3.3; see also e.g. \citealt{pat09}). Our samples are purely H$\alpha$ selected (for both cluster and field galaxies), and in this sense our cluster--field comparison would be robust. One possible bias is that our data are complete only for relatively strong emitters; we recall that our definition of star-forming galaxies is EW$>$30\AA, so that we cannot discuss faint, low-EW sources. It is likely that such low EW sources do exist in both environments (and perhaps they may be more numerous in cluster environment). However, as reported in \cite{sob11}, such low-EW sources tend to be dominated by massive galaxies with relatively low SFR (i.e.\ largely "switched-off" population), so that it would be unlikely that such low-EW sources have a significant impact on our discussions on ``star-forming'' galaxies.

\subsection{Interpretation and Caveats}

Our main finding of this study is that the SFR versus $M_*$ relation for the H$\alpha$-selected galaxies does not strongly depend on the environment at any time in the history of the Universe since $z\sim 2$ (at least when we use H$\alpha$-derived SFR). This is a similar suggestion by \cite{pen10}, who used COSMOS data to study the redshift evolution of the SSFR of (blue) star-forming galaxies, and find no environmental difference since $z\sim 1$. The independence of the SFR--$M_*$ relation with environment could be explained if the environmental quenching is a rapid process (see e.g. \citealt{muz12}). That is, the environment instantly shuts down the star-formation activity of galaxies once the environmental effects are switched on, so that declined star-formation is not observed (because our galaxy samples are selected with H$\alpha$). Therefore a naive interpretation of our result would be that the major environment quenching mechanisms are always fast-acting in the history of the Universe since $z\sim 2$.

An important, but unexplored issue is the contribution of AGNs. While most of our H$\alpha$-selected galaxies are likely to be powered by star formation (see \S~2.3), there still remains a possibility that the AGN contribution could be dependent upon redshift, mass, and environment. The ratio between H$\alpha$ and 24$\mu$m flux for AGNs can deviate more strongly than normal star-forming galaxies, depending on their dust obscuration or the observational viewing angles, which may bias the results to some extent. \cite{pop11} carried out a detailed FIR study of the star forming activity of galaxies at $z\sim 1$ using {\it  Herschel} data. They find that, while overall the SSFR--$M_*$ relation does not depend on environment, the reversal of the SFR--density relation could be produced by very massive galaxy population. They also noted that the inclusion of AGNs into the analysis could also lead to an apparent reversal of the SFR--density relation. Therefore, more detailed studies of individual galaxies (including spectroscopy) are clearly needed to unveil the role of AGNs, as a future step of this study.

Another caveat on our result concerns the prediction of dust extinction correction. We applied the empirical correction based on the $A_{\rm H\alpha}$--$M_*$ correlation established for local galaxies (see \S~2.3), which has a large intrinsic scatter (\citealt{gar10b}). The relation is reported to be unchanged out to $z\sim 1.5$ (\citealt{gar10a}; \citealt{sob12}; \citealt{iba13}), and so we do not expect the redshift evolution of the $A_{\rm H\alpha}$--$M_*$ is a major concern. However, as we showed in \S~3.3 for the $z=0.4$ galaxy sample, the dust attenuation in star-forming galaxies may be dependent upon the environment. This probably means that the ``mode'' of star formation in galaxies could be affected by the environment, leading us to underestimate the dust extinction effect of galaxies in high-density environment, if we purely rely on the $M_*$-dependent correction.

We note that the environmental dependence of ``dustiness'' of distant galaxies is still under debate. For example, \cite{pat11} used galaxies in a $z\sim 0.8$ cluster field to show that the dust extinction ($A_V$ from SED fitting) decreases with increasing galaxy number density. On the other hand, \cite{gar10a} showed that there is very little environmental variations in dust extinction ($A_{\rm H\alpha}$) by comparing IR-derived SFR with H$\alpha$-based SFRs for H$\alpha$-selected galaxy sample at a similar redshift. Our current analysis suggests an even different trend for $z=0.4$ star-forming galaxies; galaxies residing in high-density environment tend to be dustier by $\sim$0.5 mag than normal field star-forming galaxies. This may be a similar phenomenon suggested by \cite{raw12}, who find galaxies with ``warm dust'' in a $z\sim 0.3$ cluster environment using {\it Herschel} data. They suggest that these warm dust galaxies could be originated by cool dust stripping by environmental effects in cluster environments (note that the stripping preferentially removes gas from the outskirts of a galaxy). However, all these studies clearly suffer from sample size (and different definitions of star-forming galaxies and/or environment). Studying the environmental dependence of the galaxy dust properties is likely an important key for understanding the role of environment more precisely.

\section{Summary and Conclusions}

In this paper, we study the evolution and environmental dependence of the SFR--$M_*$ correlation for star-forming galaxies since $z\sim 2$. We first present the MIR properties of the H$\alpha$-selected star-forming galaxies in a rich cluster at $z=0.4$ (Cl\,0939), and then we compare the $z=0.4$ galaxies with our similar, H$\alpha$-selected galaxies at different redshifts and in different environments. Our findings are summarized as follows:

(1) The red H$\alpha$ emitters, which are reported to be most frequently seen in the group-scale environment at $z=0.4$ as shown by \cite{koy11}, are dusty red galaxies rather than passive galaxies. Using a wide-field {\it Spitzer}/MIPS 24$\mu$m dataset, we find that a large number of massive red H$\alpha$ sources are individually detected at 24$\mu$m, suggesting they are luminous and dusty. Also, with a stacking analysis, we confirm a more general trend that the red H$\alpha$ sources tend to have higher SFRs with stronger dust extinction compared with normal blue H$\alpha$ emitters.

(2) We also find that the median SFR of H$\alpha$ emitters (derived from the MIR stacking analysis) increases with increasing galaxy number density at $z=0.4$. This result is confirmed for both red and blue H$\alpha$ emitters, while the trend becomes much weaker if we compare their SSFR. We note that there still remains a positive correlation between SSFR and galaxy number density, and therefore we speculate that the SFR excess in the high-density environment can be caused by a mixed effect of both slightly higher $M_*$ and a small SSFR excess (both at $\sim$0.2~dex level) in high-density environment. This SFR increase in high-density environment amongst star-forming galaxies can (at least partially) be responsible for the reversal of the SFR--density relation claimed by recent studies.

(3) The SSFR increase towards high-density environment is {\it not} visible when we use SFRs derived from H$\alpha$ (with $M_*$-dependent extinction correction). We interpret this different trend from different SFR indicators originates from the environmental dependence of the dust attenuation for H$\alpha$ emitters. Indeed, using our $z=0.4$ sample, we find a positive correlation between $A_{\rm H\alpha}$ and galaxy number density, suggesting that star-forming galaxies ``surviving'' in high-density environment tend to be dustier than normal field galaxies (by $\sim$0.5~mag at maximum). This probably reflects a higher obscured fraction of star formation in galaxies in denser environments; e.g.\ nucleated starbursts triggered by galaxy--galaxy interactions, or the stripping effects which remove less obscured material from the outskirts of the galaxies.

(4) Using our large H$\alpha$-selected galaxy samples in distant cluster environments (from MAHALO-Subaru) and in general field environments (from HiZELS) at $z=0.4,0.8,2.2$, we examine the environmental dependence of the SFR--$M_*$ relation across cosmic time. We find that the SFR--$M_*$ relation evolves with cosmic time, but as far as we use the H$\alpha$-based SFRs, there seems to be no detectable environmental variation in the SFR--$M_*$ relation at any of these redshifts. Even if we take the possible environmental dependence of the dust extinction correction into account, we conclude that the difference in the SFR--$M_*$ sequence between cluster and field star-forming galaxies is always small ($\lsim$0.2~dex level) out to $z\sim 2$.

(5) Based on the (H$\alpha$-based) SFR--$M_*$ relation we derived for cluster and field galaxies at $z=0.4,0.8,2.2$, we also examine the evolution of the SSFR for star-forming galaxies (at the fixed mass of $M_*=10^{10}M_{\odot}$). We find that the SSFR evolves significantly, as $(1+z)^3$, in both cluster and field environments. Although the dust extinction correction applied here could be uncertain, this result suggests that the star-forming galaxy evolution in cluster environments follows the same evolutionary track as that of field galaxies. This is most simply interpreted as implying that the primary physical driver of the environmental quenching is always a fast-acting process at any time in the history of the Universe since $z\sim 2$.

In this pioneering work, we performed a comparison of the SFR--$M_*$ relation between cluster and field galaxies using the largest H$\alpha$-selected galaxy samples ever available. The most important message from this study is that the SFR--$M_*$ relation is always independent of the environment since $z\sim 2$, as far as we use H$\alpha$-based SFRs (with $M_*$-dependent extinction correction). We caution again that any environmental trend might be apparently washed out by applying the relatively simple extinction correction procedure. Future studies are clearly needed to confirm (or rule out) our finding on the ``universality'' of the SFR--$M_*$ relation across cosmic time. It may be that the ``unseen'' (obscured) star-formation activity is indeed the most important key for understanding the environmental effects across cosmic time.

\section*{Acknowledgment}
We thank the referee for their very constructive reports, which significantly improved the paper. The optical data presented in this paper are collected at the Subaru Telescope, which is operated by the National Astronomical Observatory of Japan (NAOJ). The MIR data is taken with the Spitzer Space Telescope, which is operated by the Jet Propulsion Laboratory, California Institute of Technology under a contract with NASA. Y.K. and K.I.T. acknowledge support from the Japan Society for the Promotion of Science (JSPS) through JSPS research fellowships for Young Scientists. Y.K. also acknowledges a JSPS grant allocated to NAOJ 'Institutional Programme for Young Researcher Overseas Visits' which also supported his 1-month stays at Durham (in 2010) and at MPE (in 2012). I.R.S. acknowledges support from the ERC Advanced Investigator programme DUSTYGAL 321334, a Leverhulme Fellowship, STFC and a Royal Society/Wolfson Merit Award. D.S. acknowledges the award of a Veni fellowship from the Netherlands Organisation for Scientific Research (NWO). This work was financially supported by a Grant-in-Aid for the Scientific Research (No.\, 18684004; 21340045; 24244015) by the Japanese Ministry of Education, Culture, Sports and Science.



\end{document}